\documentclass[11pt,twocolumn]{article}

\usepackage[margin=0.75in,columnsep=0.25in]{geometry}
\usepackage{amsmath,amssymb}
\usepackage[utf8]{inputenc}
\usepackage[T1]{fontenc}
\usepackage{hyperref}
\usepackage{url}
\usepackage{subcaption}
\usepackage{booktabs}
\usepackage{amsfonts}
\usepackage{nicefrac}
\usepackage{microtype}
\usepackage{xcolor}
\usepackage{graphicx}
\usepackage{textcomp}
\usepackage{enumitem}
\usepackage{authblk}
\usepackage[numbers,sort&compress]{natbib}
\usepackage{titlesec}
\usepackage{fancyhdr}
\usepackage{caption}
\usepackage{setspace}

\titleformat{\section}{\normalfont\Large\bfseries}{\thesection}{0.6em}{}
\titleformat{\subsection}{\normalfont\large\bfseries}{\thesubsection}{0.6em}{}
\titleformat{\subsubsection}{\normalfont\normalsize\bfseries}{\thesubsubsection}{0.6em}{}
\titlespacing*{\section}{0pt}{1.4em}{0.7em}
\titlespacing*{\subsection}{0pt}{1.1em}{0.5em}

\title{\Large \bfseries uMOF: A Universal Database, Benchmark, and Machine Learning Interatomic Potentials for Metal-Organic Frameworks}

\author[1,2,*,\dag]{Th\'eo Jaffrelot Inizan}
\author[1,3,\dag]{Prathami Divakar Kamath}
\author[4,*]{Alin Marin Elena}
\author[1,3,*]{Kristin A. Persson}

\affil[1]{Materials Sciences Division, Lawrence Berkeley National Laboratory, Berkeley, CA, USA}
\affil[2]{Bakar Institute of Digital Materials for the Planet, Division of Computing, Data Science, and Society, University of California, Berkeley, CA, USA}
\affil[3]{Department of Materials Science \& Engineering, University of California, Berkeley, CA, USA}
\affil[4]{Scientific Computing Department, Science and Technology Facilities Council, Daresbury Laboratory, UK}
\affil[ ]{\dag~These authors contributed equally}
\affil[ ]{*~Corresponding authors: \texttt{theo.jaf@berkeley.edu}, \texttt{alin-marin.elena@stfc.ac.uk}, \texttt{kristinpersson@berkeley.edu}}

\date{}

\begin{document}

\twocolumn[
\maketitle
\begin{abstract}
Foundation machine learning interatomic potentials (MLIPs) deliver near-ab-initio accuracy at a fraction of the computational cost, yet their promise for Metal-organic Frameworks (MOFs) remains largely unrealized as large unit cells make first-principles training data expensive to generate, fine-tuned models are scarce, and experimentally grounded benchmarks are scarcer still. We introduce uMOF, a three-part contribution addressing this gap. First, we release the largest and most accurate density functional theory (DFT) dataset for MOFs to date, computed at the r$^2$SCAN-D4 level of theory across 85{,}524 configurations spanning 19{,}950 unique frameworks and 79 elements, covering empty and gas-loaded structures, geometry optimizations, equations of state, and finite-temperature molecular dynamics (MD). Second, we release a literature-mined benchmark of 3{,}986 verified property values (3{,}146 experimental) extracted from 626 papers by a seven-stage, checkpointed multi-pass large language model (LLM) pipeline, linked to more than 650 crystallographic information files (CIFs). Third, we release two universal MLIPs for MOFs, uMOF-MH and uMOF-POLAR, fine-tuned from two architecturally distinct MACE foundation models on the uMOF dataset. On near-equilibrium, ``Tier-1'' properties (bulk modulus, phonon-derived heat capacity) the uMOF models perform comparably to existing foundation and fine-tuned baselines. On harder, dynamics-sensitive properties like gas adsorption enthalpies via Widom insertion and adsorption isotherms, the uMOF models outperform every baseline we test, including MOF-specialized gas-capture models trained on datasets up to three orders of magnitude larger, cutting error by more than 80\% to within experimental uncertainty. We trace this advantage to the physical diversity of the training data and to level of theory where a small (1.7\%) fraction of MD simulations is decisive for MLIP stability, and r$^2$SCAN-D4 captures guest-MOF and metal-linker interactions that GGA-level (PBE) datasets miss. We release the dataset, benchmark, and models as a reproducible standard for MLIP development in reticular chemistry.
\end{abstract}
\vspace{1.2em}
]
\thispagestyle{fancy}

\section{Introduction}
\label{sec:intro}

Foundation machine-learning interatomic potentials (MLIPs) \citep{wood2026family,neumann2024orb,deng2023chgnet,batatia2025foundation,chen2022universal} have transformed atomistic simulation over the recent years, replacing expensive quantum-mechanical energy and force evaluations with learning surrogates that approach denisty functional theory (DFT) accuracy at orders of magnitude lower cost. Trained across a substantial fraction of the periodic table on thousands to millions of DFT data points \citep{mptrj,barros2026open,ghahremanpour2018alexandria,kaplan2025matpes}, these models promise transferability from a single pretrained checkpoint \citep{batatia2022mace,merchant2023gnome,batatia2024foundation,riebesell2024matbench}, driving rapid adoption for crystal-structure prediction, high-throughput screening, and materials discovery. Equivariant message-passing backbones such as MACE \citep{batatia2022mace} are among the most widely used, and while early foundation models were trained at the PBE or PBE+U level, more recent efforts have moved toward meta-GGA functionals \citep{kaplan2025matpes,kuner2025mp} such as r$^2$SCAN \citep{r2scan}, which reduce self-interaction and delocalization error relative to PBE \citep{kaplan2025matpes,batatia2024foundation}. Despite this progress, training corpora remain dominated by dense inorganic bulk crystals and small molecules, leaving structurally distinct material classes underrepresented and their MLIP accuracy largely unevaluated.

Metal-organic frameworks (MOFs) \citep{mof,mof1,mof2} are a particularly demanding test case as a MOF unit cell couples inorganic metal nodes, organic linkers, and often guest molecules within porous, low-density lattices exceeding a thousand atoms per cell. The same chemistry that makes MOFs valuable for gas storage \citep{li2018recent}, separation \citep{li2012metal}, and catalysis \citep{lee2009metal} also makes them combinatorially diverse and physically distinct from the training distribution of most foundation models \citep{kamath,mofsim}. Properties central to MOF applications like adsorption enthalpies, thermal expansion, framework flexibility are governed by long-range dispersion interactions \citep{oliveira2026flames,r2scanmof} and finite-temperature dynamics that short-range, near-equilibrium-trained foundation models are not designed to capture. Compounding this, because large unit cells make DFT single-point calculations expensive, MOF training data \cite{kamath,opendac,sriram2025open,lim2025accelerating} is built almost exclusively at the PBE \citep{pbegga} level, with dispersion corrections applied inconsistently or omitted, even though PBE is known to poorly capture spin-polarization in open-shell metal nodes \citep{u} and the van der Waals interactions governing guest binding \citep{rehak2020including, vlaisavljevich2017performance}.

 Current MOF-specific MLIPs fall majorly into three families: active-learning, system-specific models \citep{eckhoff2019mof5,vandenhaute2023incremental,wieser2024mlff,sharma2024tempactive} that do not scale across chemical space; guest-adsorption-focused models \citep{goeminne2023dftquality,liu2024h2oms,sriram2024odac,lim2025accelerating}, which inherit PBE's systematic mis-description of dispersion-dominated guest-framework interactions relative to higher-level functionals \citep{cho2024improving,vlaisavljevich2017performance}; and foundation-model fine-tunes like MACE-MP-MOF0 \citep{kamath}, a PBE+D3(BJ)\citep{d3,bj} fine-tune restricted to 24 elements and developed mainly for near-equilibrium properties, leaving guest-adsorption and high-temperature behavior insufficiently captured. On the benchmarking side, general-purpose suites like Matbench Discovery \citep{riebesell2024matbench,chiang2026mlip} offer little MOF-specific coverage and primarily compare against DFT rather than curated experimental data at scale. LLM-based literature mining \citep{kang2025harnessing} offers promise but so far is an underexplored route to scalable, experimentally grounded benchmarks.

This paper introduces uMOF, a dataset, benchmark, and model release designed to close these gaps simultaneously. We contribute: (i) the uMOF dataset (\S\ref{sec:dataset}), the first r$^2$SCAN-D4 \citep{caldeweyher2019generally} DFT dataset for MOFs at scale with 85{,}524 configurations from 19{,}950 frameworks spanning 79 elements, combining finite-temperature MD, relaxation/equation-of-state trajectories, and Gaussian noise perturbed structures, plus six gas adsorbates in industrially relevant empty and loaded MOFs, (ii) a literature-mined experimental benchmark (\S\ref{sec:benchmark}), 3{,}986 verified property records (3{,}146 experimental) across 20 property classes extracted from 626 papers via a multi-stage LLM workflow with deterministic hallucination filtering, linked to over 650 CIF structures and released as an open module of \texttt{ml-peg} \citep{mlpeg}, (iii) the uMOF models (\S\ref{sec:models}), uMOF-MH and uMOF-POLAR, two architecturally distinct MACE \citep{batatia2022mace} fine-tunes that substantially improve on their parent foundation models and on MACE-MP-MOF0 \citep{kamath} on gas-adsorption properties; and (iv) a fine-tuning and data-generation protocol (\S\ref{sec:ablation}) showing that finite-temperature MD, even at $\sim$1.7\% of the training set, is decisive for off-equilibrium force-model stability, and that level of theory matters more than raw data volume: uMOF outperforms gas-capture-specialized models like UMA-ODAC trained on $\sim$1{,}000$\times$ more data by over 80\% on adsorption enthalpy.

\begin{figure*}[t]
    \centering
    \includegraphics[width=0.85\linewidth]{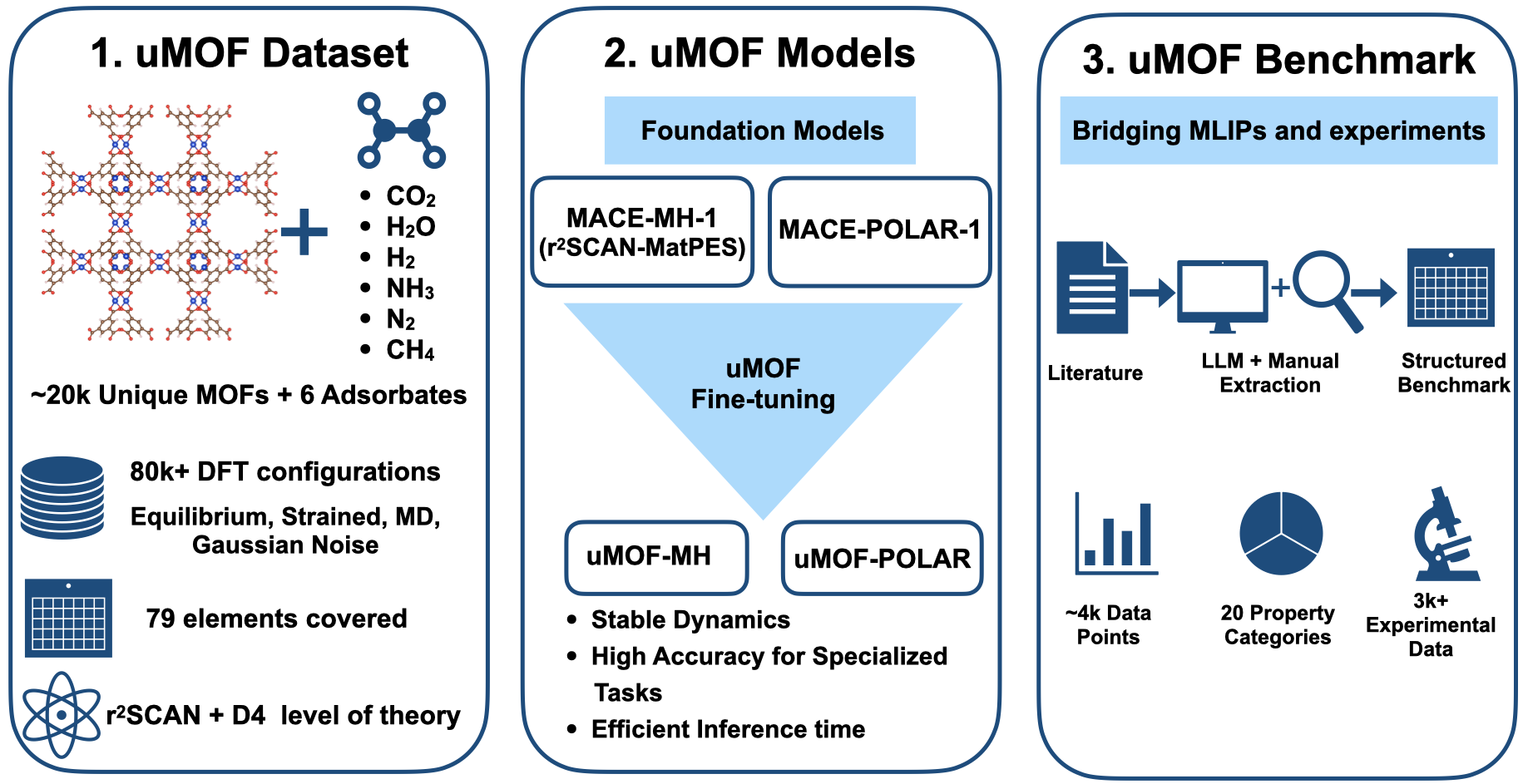}
    \caption{\textbf{Overview of uMOF}: a r$^2$SCAN-D4 DFT dataset, two fine-tuned foundation MLIPs, and a literature-mined experimental benchmark that together bridge MLIP development and MOF experiments.}
    \label{fig:umof}
\end{figure*}

\section{Methods}
\subsection{The uMOF Dataset}
\label{sec:dataset}

\subsubsection{Level of theory}
\label{sec:dataset_theory}
Every configuration in the uMOF dataset carries DFT energies, forces, and stresses computed at the r$^2$SCAN-D4 level: the r$^2$SCAN meta-GGA functional \citep{furness2020r2scan} improves systematically on PBE, while remaining tractable at the scale required for an 85{,}524-configuration dataset. The charge-dependent D4 dispersion correction \citep{caldeweyher2019d4} is applied throughout and, as we show quantitatively in Appendix~\ref{sec:appendix_dft}, benchmarking optimized geometries against experiment for a non-spin-polarized (Zn-MOF-5) and a spin-polarized (Cu-HKUST-1) framework shows r$^2$SCAN-D4 gives 2--4\% lower errors than PBE-D3(BJ) (Appendix Figure \ref{fig:r2scan}). These framework-level gains are consistent with broad molecular benchmarks, where r$^2$SCAN-D4 ranks among the most accurate meta-GGAs across thermochemistry, transition-metal and metal-organic reactions \citep{dohm2018mor41}, host-guest non-covalent interactions, and dispersion-bound molecular-crystal energies and equations of state \citep{ehlert2021r2scand4, dohm2018mor41}.

All DFT calculations were performed with VASP 6.4.2 \citep{vasp, vasp1, vasp2, vasp3} using the \texttt{atomate2} workflow manager \citep{ganose2025atomate2}. Plane-wave cutoff energy (ENCUT) was selected by benchmarking 520, 680, 720, and 900\,eV for cost versus accuracy in reproducing experimental unit-cell parameters and bulk moduli; we settled on ENCUT~$=680$\,eV with EDIFF~$=10^{-5}$\,eV and \texttt{LREAL=False} for accurate forces. Spin-polarization was turned on (ISPIN=2) only for open-shell-metal-based MOFs as r$^2$SCAN+D4 scales expensively with spin-polarization for large systems like MOFs. Remaining INCAR settings follow the MatPES r$^2$SCAN static-calculation generator (\texttt{MatPesMetaGGAStaticSetGenerator}) in \texttt{atomate2}, consistent with the broader MatPES effort \citep{kaplan2025matpes}.

\subsubsection{Sampling strategy and dataset composition}
\label{sec:dataset_sampling}
A dataset intended to train a transferable, dynamics-capable MOF potential must cover more than relaxed, near-equilibrium geometries. We combine three complementary sampling strategies (Table~\ref{tab:data}): (i) Gaussian-noise perturbation seeded from QMOF structures \citep{qmof1, qmof2}, which cheaply covers a broad range of chemistries at modest structural displacement ($\sigma \approx 0.06$\,\AA); (ii) geometry-relaxation and equation-of-state (EOS) trajectories, which cover the near-equilibrium potential-energy surface relevant to elastic and thermodynamic properties; and (iii) finite-temperature molecular dynamics, which is the only source of genuinely off-equilibrium, high-force configurations in the dataset spanning a wide range (mean $|F| = 2.48$\,eV\,\AA$^{-1}$, 95th percentile $7.60$\,eV\,\AA$^{-1}$). These are supplemented with curated industrially relevant empty and adsorbate-loaded frameworks across six gas species, and isolated-atom and molecule references required for consistent energy referencing during training.

In total, the dataset spans 85{,}524 configurations across 19{,}950 unique MOFs and 79 elements ($Z \in \{1, 3\text{--}9, 11\text{--}17, 19\text{--}35, 37\text{--}53, 55\text{--}83, 90, 92\text{--}94\}$). The same framework can appear in more than one sampling category (e.g., both as a Gaussian-noise structure and as an MD trajectory seed), so the totals in Table \ref{tab:data} are distinct counts rather than column sums. 

\begin{table*}[t]
\small
\caption{\textbf{Composition of the uMOF dataset}. The \emph{Molecular dynamics} row aggregates three trajectory sets: Cu-MOF MD (1{,}433 frames, $\sim$70 Cu-based frameworks), temperature-sampled MOF MD (4{,}062 frames, $\sim$160 chemically varied frameworks spanning In, Zn, Mg, Al, and predominantly other closed-shell metal MOFs), and empty-framework MD (492 frames, 6 guest-free frameworks).}
\label{tab:data}
\centering
\begin{tabular}{lrr}
\toprule
Data Type & Unique Count & Total Configurations \\
\midrule
Gaussian noise                      & 19{,}922 MOFs & 59{,}661 \\
Geometry optimization                & 1{,}288 MOFs  & 12{,}621 \\
Molecular dynamics                   & 236 MOFs      & 5{,}987 \\
Strained, equation of state          & 143 MOFs      & 810 \\
Adsorbate-loaded frameworks          & 13 MOFs $\times$ 6 adsorbates & 5{,}646 \\
Isolated atoms                       & 79 elements   & 79 \\
Isolated gas molecules               & 6 adsorbates  & 720 \\
\midrule
\textbf{Total (distinct)}            & \textbf{19{,}950} & \textbf{85{,}524} \\
\bottomrule
\end{tabular}
\end{table*}

\subsubsection{Training/validation/test splits.} The core training corpus from uMOF (\texttt{h5\_mof}) contains 68{,}394 train, 8{,}470 validation and 8{,}660 test frames (6{,}823{,}332 / 847{,}345 / 859{,}897 atoms respectively), drawn from 21{,}519 distinct frameworks: 21{,}427 QMOF structures \citep{rosen2021qmof} plus roughly 90 common-name experimental MOFs (including UiO, MIL, ZIF, HKUST, MOF-74, IRMOF, CAU, and CALF-20 families). A separate Gaussian-noise augmentation set (\texttt{h5\_mofgen}), generated from structures produced by our MOFGen diffusion model \citep{inizan2025agenticaidiscoverymetalorganic}, adds 37{,}732 train and 4{,}192 validation frames (no held-out test split), consisting of DFT relaxation trajectories ($\approx$12.5 frames per framework) for 3{,}350 generated frameworks disjoint from QMOF. Combined, the two sources total 127{,}448 configurations (12{,}885{,}556 atoms) across 24{,}869 unique frameworks.  A neighbor cutoff of $r_{\max} = 4.5$\,\AA{} yields approximately 30 neighbors per atom on average across the dataset. As discussed in \S\ref{sec:models}, we ultimately found that including the diffusion-generated augmentation set at training time degraded downstream accuracy, so the final released models are fine-tuned on the uMOF \texttt{h5\_mof} alone (Table~\ref{tab:data}), while \texttt{h5\_mofgen} is released alongside it for the community to use as they see fit.

\subsection{The uMOF Benchmark}
\label{sec:benchmark}

Assessing whether an MLIP is actually useful for MOF applications requires comparison against experiment, not just against DFT. But experimental MOF property measurements are scattered across thousands of papers, reported in inconsistent units, under varying conditions, and often only as figure data rather than tabulated values. Manually curating such a benchmark at scale is prohibitively labor-intensive. We instead build an automated multi-pass LLM pipeline to extract property values directly from manually downloaded papers, with deterministic verification steps designed to suppress hallucination.

\subsubsection{Pipeline design}
\label{sec:benchmark_pipeline}
The benchmark is produced by a multi-stage pipeline operating over MOF and COF papers, where each stage reads and writes a versioned JSON file so that any individual stage can be re-run or audited independently of the others.

\begin{enumerate}[leftmargin=1.6em,itemsep=1pt,topsep=2pt]
\item \textbf{Load.} Collect articles and candidate DOIs for which machine-readable text chunks are available.
\item \textbf{Rank} (Claude Haiku~4.5). Score each paper's title and abstract on a 0 to 5 relevance scale for simulation-relevant content, processed in batches of 20 papers.
\item \textbf{Extract} (Claude Opus~4.7). For papers scoring above the threshold, emit one record per extracted property: a value, unit, measurement conditions (temperature, pressure, guest species), level of theory or experimental method, source location within the paper, a verbatim supporting quote with chunk index, a flag for whether the value is figure-only, and a reproducibility verdict covering feasibility, computational backend, workflow, and MLIP-compatibility.
\item \textbf{Verify.} Deterministically fuzzy-match each extracted quote against its source chunk within a $\pm2$-chunk window, requiring Levenshtein similarity $\ge 0.95$; this stage makes no further LLM calls, so it cannot itself introduce new hallucinations. 
\item \textbf{Score and filter.} Aggregate a per-paper reproducibility score from the verified-quote fraction, presence of a code or CIF reference, and completeness of the reported method; drop properties flagged infeasible and drop entire papers with verified fraction below 0.6 or with no viable route to a CIF structure. The reproducibility score is a proxy for, not a guarantee of, correctness. 
\item \textbf{Multimodal upgrade} (Claude Opus~4.7). Re-read figure-locked values directly from rendered figure images for up to 50 papers, recovering values that are visually but not textually reported.
\item \textbf{Export.} Assemble the final \texttt{benchmark.json} file with manually added advanced entries like inelastic neutron scattering (INS) spectra compiled from literature review, together with a per-stage audit file recording record counts, drop reasons, API cost, model versions, configuration hash, and git commit SHA for full reproducibility. 
\end{enumerate}

\subsubsection{Property tiers and dataset statistics}
\label{sec:benchmark_stats}
We organize the resulting 20 property classes into three tiers reflecting the cost of reproducing them with an MLIP calculator: \textbf{Tier 1} covers static or finite-difference properties (lattice parameters, cohesive/formation energy, elastic tensor, bulk/shear/Young's modulus, equation of state, heat capacity, INS spectra); \textbf{Tier 2} covers short ($\le 1$\,ns) molecular dynamics properties (NPT density, thermal expansion, self-diffusion, radial distribution function, NVT stability); and \textbf{Tier 3} covers specialized-but-inexpensive properties centered on adsorption (Henry coefficients via Widom insertion, isosteric heat of adsorption at low loading, binding-site enumeration). The full breakdown is given in Table~\ref{tab:bench}.

The pipeline processed 626 candidate papers into 3{,}986 verified property records (79\% experimental, 3{,}146 records) linked to more than 650 CIF structures drawn from CoRE-MOF~2025 \citep{chung2019coremof} and MOSAEC-DB \citep{gibaldi2025mosaec}. Structures are linked to their source papers by DOI 272 of 610 source papers (44.6\%) resolve to a deposited CIF, yielding 830 candidate CIFs total.

\begin{table*}[t]
\small
\caption{The uMOF benchmark by property class. Tier 1 = static/finite-difference, Tier 2 = short MD, Tier 3 = specialized (adsorption-centric).}
\label{tab:bench}
\centering
\begin{tabular}{lrrr}
\toprule
Property Name & Tier & No.\ of data points & No.\ of experimental data points \\
\midrule
Lattice parameters        & 1 & 2913 & 2656 \\
$Q_{st}$ (low loading)     & 3 & 283  & 225 \\
Binding energy             & 1 & 222  & 37 \\
Binding-site enumeration   & 3 & 104  & 59 \\
MSD / self-diffusion       & 2 & 71   & 5 \\
Bulk modulus               & 1 & 69   & 20 \\
NPT density                & 2 & 61   & 55 \\
Cohesive energy            & 1 & 60   & 0 \\
Henry coefficient          & 3 & 59   & 31 \\
Thermal expansion          & 2 & 37   & 17 \\
Formation energy           & 1 & 25   & 0 \\
Young's modulus            & 1 & 15   & 13 \\
Elastic tensor             & 1 & 14   & 0 \\
INS spectrum               & 1 & 12   & 7 \\
Shear modulus              & 1 & 10   & 0 \\
Cell vs.\ pressure         & 1 & 8    & 8 \\
Heat capacity              & 1 & 7    & 7 \\
Equation of state          & 1 & 6    & 3 \\
Radial distribution fn.\   & 2 & 6    & 1 \\
NVT stability              & 2 & 4    & 2 \\
\midrule
\textbf{Total} & & \textbf{3{,}986} & \textbf{3{,}146} \\
\bottomrule
\end{tabular}
\end{table*}

The benchmark will also be released as an open module of \texttt{ml-peg} \citep{mlpeg}\footnote{\url{https://github.com/ddmms/ml-peg}}, independent of the specific models introduced in this paper, so that any MLIP can be scored against curated MOF experiments in a reproducible way.

\section{Models and Fine-Tuning}
\label{sec:models}

We fine-tune two architecturally distinct MACE \citep{batatia2022mace} foundation models on the uMOF dataset, chosen to probe whether an explicit long-range electrostatic term is beneficial for MOF chemistry beyond what a well-chosen dispersion correction and a higher level of theory already provide. \textbf{(i) uMOF-MH} fine-tunes the r$^2$SCAN-D4 MatPES head of MACE-MH-1, a multi-head foundation model with a short-range cutoff and no explicit long-range electrostatic term \citep{batatia2024foundation}. \textbf{(ii) uMOF-POLAR} fine-tunes MACE-POLAR-1-M, a polarizable variant of MACE that includes an explicit long-range electrostatic term conditioned on total charge and spin multiplicity \citep{grisafi2019longrange}, with elemental coverage up to $Z \le 83$.

Both models use isolated-atom energy references, a two-stage training schedule with stochastic weight averaging (SWA), and a force/stress-weighted loss in the second stage. uMOF-MH and uMOF-POLAR additionally retains a replay head from their foundation training to limit catastrophic forgetting of the broader chemical space seen during pretraining. For comparison against an existing PBE-level baseline, we also evaluate MACE-MP-MOF0 (v2) \citep{kamath}, a PBE+D3(BJ) MOF fine-tune of MACE-MP-0b, allowing us to isolate the effects of base architecture, training data, and level of theory on downstream accuracy. Table \ref{tab:models_appendix} summarizes the model panel and held-out training accuracy: both uMOF fine-tunes converge to small held-out force errors ($\lesssim 0.2$\,eV\,\AA$^{-1}$ RMSE), confirming that fine-tuning successfully transfers the foundation representation to r$^2$SCAN-D4 without affecting it.

\begin{table*}[t]
\small
\caption{\textbf{Model panel and training accuracy.} Held-out validation errors (best epoch, r$^2$SCAN-D4 head, per-atom energy) are reported for the two uMOF fine-tunes}
\label{tab:models_appendix}
\centering
\begin{tabular}{lllccc}
\toprule
Model & Base model & Training data/head & Long-range & RMSE-$E$ & RMSE-$F$ \\
& & & & (meV/atom) & (eV\,\AA$^{-1}$) \\
\midrule
MACE-MH-1+D3      & foundation      & MatPES (r$^2$SCAN)          & add-on    & --- & --- \\
\textbf{uMOF-MH}   & MACE-MH-1       & uMOF (r$^2$SCAN-D4)          & from DFT  & 33.2 & 0.066 \\
MACE-POLAR-1-M     & foundation      & OMol ($\omega$B97M-V-rv10)   & yes       & --- & --- \\
\textbf{uMOF-POLAR}& MACE-POLAR-1-M  & uMOF (r$^2$SCAN-D4)          & yes       & 39.4 & 0.211 \\
MACE-MP-MOF0-v2    & MACE-MP-0b      & MOF0 (PBE+D3(BJ))            & from DFT  & --- & --- \\
\bottomrule
\end{tabular}
\end{table*}

\section{Results and Discussion}
\label{sec:results}

\subsection{Molecular dynamics data is decisive for training stability}
\label{sec:ablation}

The molecular dynamics portion of the uMOF dataset (5{,}987 frames total, Table~\ref{tab:data}) aggregates three trajectory sets: Cu-MOF MD set (1{,}433 frames across roughly 70 Cu-based frameworks as Cu is an abundant and open-shell metal that is challenging to model in MOFs), other temperature-sampled MOF MD (4{,}062 frames across roughly 160 chemically varied, predominantly closed-shell-metal frameworks spanning Zr, Zn, Mg, and Al), and empty-framework MD (492 frames across 6 guest-free frameworks). To isolate the contribution of finite-temperature sampling, we removed only the Cu-MOF MD subset (1{,}146 training frames, 1.7\% of the training set) from an otherwise identical MACE-MH-1 fine-tuning recipe, holding validation and test splits fixed.

Even this small removal has an outsized effect (Table~\ref{tab:ablation}). The two runs are essentially identical through the first training stage but once the force/stress-weighted SWA stage begins, the reduced-data model's validation force error spikes to 0.154\,eV\,\AA$^{-1}$ and settles roughly 25\% worse than the full-data model (0.101 vs.\ 0.081\,eV\,\AA$^{-1}$ final RMSE-$F$), while energy accuracy is essentially unchanged (+1.2\,meV/atom).

\begin{table}[t]
\small
\caption{MD ablation (identical MACE-MH-1 fine-tune recipe). Removing the Cu-MOF MD subset (1{,}146 frames, 1.7\% of the training set) leaves pre-SWA force accuracy unchanged and energy essentially unchanged (+1.2\,meV/atom), but degrades the force-weighted SWA stage by $\sim$25\%.}
\label{tab:ablation}
\centering
\resizebox{\linewidth}{!}{%
\begin{tabular}{lcccc}
\toprule
Training data & Frames & RMSE-$E$ [meV/atom] & RMSE-$F$ pre-SWA [eV\,\AA$^{-1}$] & RMSE-$F$ final [eV\,\AA$^{-1}$] \\
\midrule
With Cu-MOF MD    & 68{,}394 & 37.3 & 0.081 & \textbf{0.081} \\
Without Cu-MOF MD & 67{,}248 & 38.5 & 0.081 & 0.101 \\
\bottomrule
\end{tabular}}
\end{table}

 We re-evaluated both stage-two models on a per-configuration-type basis across the full validation set (Table~\ref{tab:ablation_pertype}). The degradation is not confined to the held-out MD frames: the QMOF Gaussian-noise single-point structures, which make up 70\% of the validation set and are chemically distinct from the removed Cu-based frameworks, degrade by the same $\sim$25\%. Near-equilibrium categories, geometry optimization, equation of state, MOF0-derived structures, adsorbate-loaded frameworks, and empty frameworks barely move (+3--4\%). Both finite-temperature MD snapshots and Gaussian-noise-perturbed statics populate the off-equilibrium, high-force regime of the potential-energy surface and the Cu-MOF MD data teaches the model exactly this regime. The force-weighted SWA stage relies on that signal precisely where the bulk of the QMOF Gaussian-noise data lives. Molecular dynamics is the only part of the corpus that samples the high-force regime governing dynamics, Widom insertion, and anharmonic phonon calculations downstream, so even a 1.7\% Cu-MOF MD slice functions as a decisive, non-optional ingredient of MOF fine-tuning data rather than a minor augmentation.

\begin{table*}[t]
\small
\caption{Per-configuration-type validation force RMSE (eV\,\AA$^{-1}$) for the MD ablation of Table~\ref{tab:ablation}. Removing the Cu-MOF MD subset degrades not only the held-out MD frames but the chemically distinct QMOF Gaussian-noise structures (70\% of validation) by the same $+25\%$, while near-equilibrium categories barely change -- evidence that finite-temperature sampling transfers to the broader off-equilibrium force landscape.}
\label{tab:ablation_pertype}
\centering
\begin{tabular}{lrccr}
\toprule
Configuration type & $n$ (val) & With Cu-MOF MD & Without & $\Delta$ \\
\midrule
QMOF Gaussian noise (single-point) & 5{,}967 & 0.090 & 0.113 & $+25\%$ \\
MD (Cu-MOF)                        & 143     & 0.074 & 0.112 & $+51\%$ \\
Equation of state                  & 116     & 0.065 & 0.068 & $+4\%$ \\
Geometry optimization               & 1{,}253  & 0.054 & 0.055 & $+3\%$ \\
MOF0-derived                        & 301     & 0.040 & 0.041 & $+4\%$ \\
Adsorbate-loaded                    & 547     & 0.025 & 0.026 & $+4\%$ \\
Empty frameworks                    & 71      & 0.020 & 0.021 & $+4\%$ \\
Isolated molecules                  & 72      & 0.012 & 0.021 & $+73\%$ \\
\midrule
\textbf{All}                        & \textbf{8{,}470} & \textbf{0.081} & \textbf{0.101} & \textbf{$+25\%$} \\
\bottomrule
\end{tabular}
\end{table*}

\subsection{Preserving accuracy on near-equilibrium mechanical and thermal properties}
\label{sec:tier1}

We first evaluate near-equilibrium, Tier-1 properties across nine benchmark MOFs spanning seven metal nodes and diverse linker chemistries (structures and experimental references listed in Appendix~\ref{sec:bench}). MACE-MH-1 \emph{without} any dispersion correction achieves the lowest bulk-modulus error of any model tested (Table~\ref{tab:tier1}), consistent with the tendency of D3/D4-type corrections to overestimate long-range interactions in MOFs \citep{longrange}. MACE-MP-MOF0, which is fine-tuned on a smaller, more targeted PBE+D3(BJ) dataset that overlaps heavily with the benchmarked structures, performs comparably. The uMOF models, fine-tuned on a dataset two orders of magnitude larger and spanning 79 vs.\ 24 elements, trade at most 0.04\,GPa of bulk-modulus accuracy for substantially broader elemental and structural coverage. On this metric taken alone, model choice looks largely unaffected.

Phonon-derived heat capacity ($C_v$), computed alongside the fraction of spurious imaginary phonon modes, tells a more cautionary story. MACE-MH-1 without dispersion again appears the most accurate model on aggregate $C_v$ error, but this is an artifact as it produces the largest fraction of spurious imaginary modes, which are discarded before computing $C_v$ and thereby silently remove exactly the contributions that would otherwise drive the metric toward larger, more realistic errors. This rewards an unstable potential and penalizes MACE-MP-MOF0, the only model in our panel with 0.00\% imaginary modes with the largest apparent $C_v$ error, precisely because it alone produces a physically complete vibrational spectrum within the defined convergence criteria (Appendix \ref{sec:appendix_params}. Comparison INS spectra, an independent experimental observable, corroborates this reading where MACE-MH-1's low $C_v$ error coexists with one of the largest INS Wasserstein distances from experiment due to the missed spurious imaginary modes(Appendix Figure~\ref{fig:ins}), confirming that its apparent $C_v$ accuracy is an artifact of discarded modes rather than a genuinely correct vibrational spectrum. This echoes and independently corroborates the findings in \citep{kamath2026impact} that aggregate accuracy metrics can reward models for the right answer for the wrong reason unless imaginary-mode fraction is explicitly audited; we recommend this auditing as standard practice for MLIP vibrational-property benchmarking going forward.

\begin{table*}[t]
\footnotesize
\caption{\textbf{Tier-1 property accuracy.} Bulk-modulus MAE (nine MOFs); phonon-derived $C_v$ error at 300\,K, \% imaginary modes, and INS Wasserstein distance at 10\,K (seven MOFs).}
\label{tab:tier1}
\centering
\begin{tabular}{lcccc}
\toprule
Model & Bulk modulus MAE [GPa] & MPAE$_{C_v}$ (\%) & \% Imaginary modes & Wasserstein dist.\ [meV] \\
\midrule
MACE-MH-1        & 3.54 & 7.22 & 0.39 & 3.25 \\
\textbf{uMOF-MH}   & 3.92 & 8.59 & 0.25 & 3.18 \\
MACE-POLAR-1-M   & 4.69 & 7.88 & 0.38 & 2.86 \\
\textbf{uMOF-POLAR}& 4.94 & 7.59 & 0.28 & 3.27 \\
MACE-MP-MOF0-v2  & 3.88 & 9.72 & 0.00 & 2.98 \\
\bottomrule
\end{tabular}
\end{table*}

Overall, Tier-1 analysis indicates that the uMOF models achieve accuracy comparable to existing foundation and fine-tuned baselines on generic, near-equilibrium properties.

\begin{figure*}[t]
    \centering
    \includegraphics[width=0.85\linewidth]{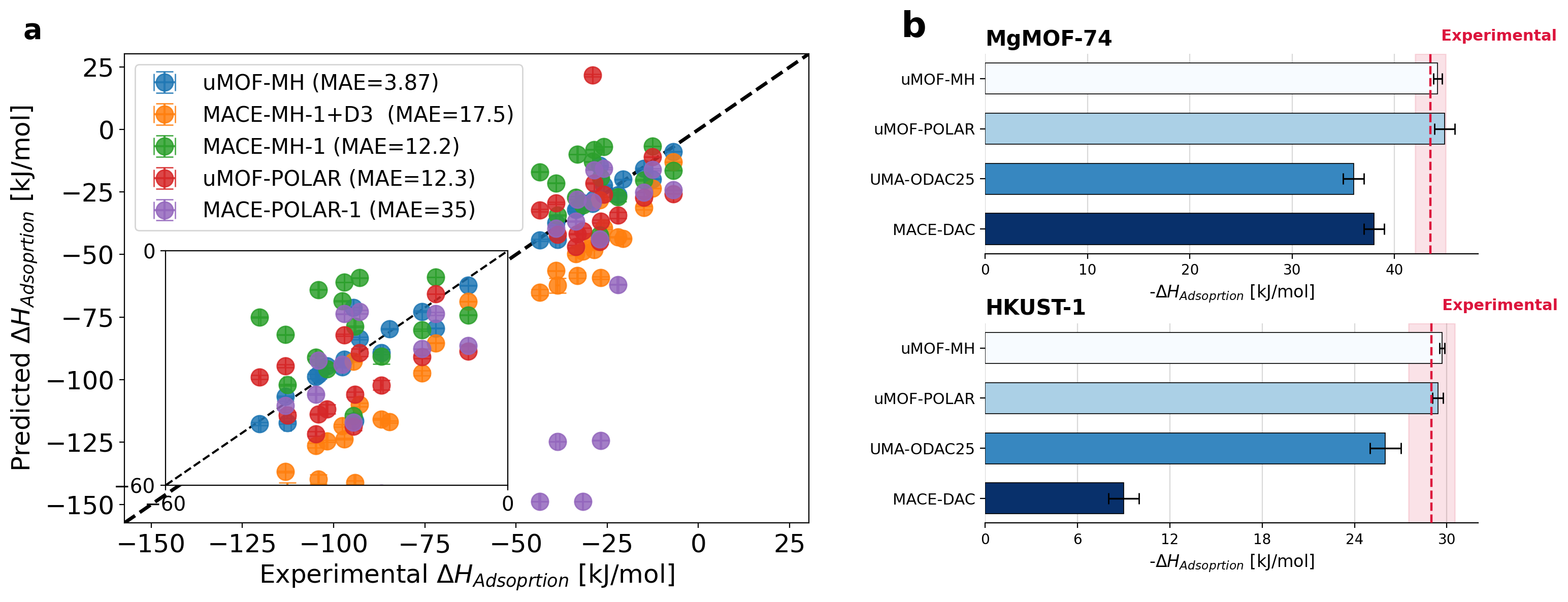}
    \caption{\textbf{Model accuracy on experimental binding enthalpies from Widom insertion:} (a) uMOF-MH and uMOF-POLAR models outperform their pre-trained foundation models by more than 2 times. (b) The uMOF models show a clear advantage, with errors improved by more than 80\% over other gas-capture-focused models for MOFs such as UMA-ODAC25 and MACE-DAC. Values for UMA-ODAC25 and MACE-DAC were extracted from \citet{oliveira2026flames}, plotted with a standard deviation of 1\,kJ/mol. Experimental values are taken from \citet{queen2014comprehensive} and \citet{grajciar2011understanding}. }
    \label{fig:adsorption_site_enthalpy}
\end{figure*}
\subsection{Gas adsorption: where uMOF models clearly pull ahead}
\label{sec:adsorption}

Tier-1 tasks understate what matters for real MOF applications. Gas adsorption is one of the most widely studied MOF applications, and accurately predicting adsorption enthalpies and binding sites remains a central, long-standing open challenge for MLIPs \citep{kancharlapalli2023high}. Unlike bulk modulus and harmonic phonons, which probe the near-equilibrium potential-energy surface, adsorption requires jointly and correctly capturing both framework dynamics and guest-host dispersion interactions which is exactly the regime that \S\ref{sec:ablation} shows is governed by the MD fraction of the training data.

Using the \texttt{flames} package \citep{oliveira2026flames}, we run Widom-insertion calculations at 298\,K for 20{,}000 steps (convergence checked against step count on HKUST-1 as a representative case; full parameters in Appendix~\ref{sec:appendix_params}) across 13 MOFs for CO$_2$ and compare predicted binding enthalpies against experiment. We further test adsorption of N$_2$, CH$_4$, and H$_2$ to test adsorbate sensitivity and accuracy in HKUST-1, a challenging test case. MACE-MP-MOF0 is excluded from this comparison, as it was fine-tuned specifically for phonon accuracy and is not suited to gas-adsorption calculations. Instead, other domain-specific models like UMA-ODAC \cite{wood2026family,sriram2025open}and MACE-DAC \citep{lim2025accelerating} trained specifically for MOF gas-adsorption applications are used for comparison.

The uMOF models are clearly and consistently superior to their corresponding foundation-model baseline (Figure~\ref{fig:adsorption_site_enthalpy} a), more than halving error to a mean absolute error (MAE) of 4\,kJ/mol effectively reproducing experimental values considering experimental uncertainty for these systems \citep{park2017reproducible}. Dispersion treatment matters throughout for MOFs where MACE-MH-1 without any dispersion correction significantly underestimates adsorption enthalpy, while adding D3 leads to significant overestimation, so neither variant achieves a good balance. MACE-POLAR-1, despite including an explicit long-range electrostatic term, performs worst of all baselines tested, a counter-intuitive result. Fine-tuning on uMOF's r$^2$SCAN-D4 data resolves this balance for both architectures, indicating that the gain comes from the level of theory and the composition of the training data rather than from architectural choices. The uMOF models surpass domain-specialized models trained explicitly for gas-capture applications by more than 80\% on adsorption enthalpy MAE. UMA-ODAC25 is trained on ODAC2025, a PBE+D3 dataset of roughly 70 million MOF-adsorbate configurations which is nearly three orders of magnitude larger than uMOF's training corpus. MACE-DAC is trained on the smaller, related GoldDAC dataset. That a dataset almost $1{,}000\times$ smaller than ODAC2025 outperforms models purpose-built for gas capture at that scale is a direct demonstration that data volume alone does not guarantee downstream accuracy for this class of properties and level of theory matters at least as much, if not more. uMOF's r$^2$SCAN-D4 labels move beyond the GGA-level PBE used throughout existing MOF datasets, and this accuracy advantage persists despite uMOF's relatively far smaller scale.

\subsubsection{Binding-site characteristics.} Accuracy on adsorption enthalpy is not achieved by chance as we further evaluate binding-site characteristics across the M-MOF-74 series (M~=~Mg, Fe, Co, Ni, Mn, Cu, Zn), where CO$_2$ is experimentally known to bind at the open metal site \citep{wasik2024impact}. uMOF-MH identifies the correct open-metal binding site as the most stable for 4 of 7 metals (Mg, Ni, Mn, Zn); for M~=~Fe, Co, and Cu, the model instead places the molecule in the pore. For the correctly identified sites, the most accurate uMOF-MH reproduces the experimental M--O(CO$_2$) bond distance and M--O--C bond angle to within 0.06\,\AA{} and 4.5$^\circ$ on average (Table~\ref{tab:M-MOF-74}, and the model reproduces the experimental trend of adsorption strength (Mg > Ni > Co > Fe > Mn > Cu,) across this metal-substitution series . Zn-MOF-74 is the only variant for which the uMOF-Mh model fails to get the right trend.

\begin{table}[t]
\small
\caption{uMOF-MH errors relative to experiment for M--O(CO$_2$) bond distance and M--O--C bond angle across correctly identified binding sites in the M-MOF-74 series (M = Mg, Ni, Mn, Zn). Experimental references from \citet{queen2014comprehensive}.}
\label{tab:M-MOF-74}
\centering
\begin{tabular}{lrr}
\toprule
Metal (M) & $|\Delta_{\text{distance}}|$ [\AA] & $|\Delta_{\text{angle}}|$ [$^\circ$] \\
\midrule
Mg & 0.11  & 4.2 \\
Ni & 0.03  & 3.5 \\
Mn & 0.005 & 3.2 \\
Zn & 0.101 & 7.18 \\
\midrule
\textbf{MAE} & \textbf{0.06} & \textbf{4.52} \\
\bottomrule
\end{tabular}
\end{table}

\subsubsection{Adsorption isotherms.} 
\begin{figure}[t]
\centering
\includegraphics[width=1.0\linewidth]{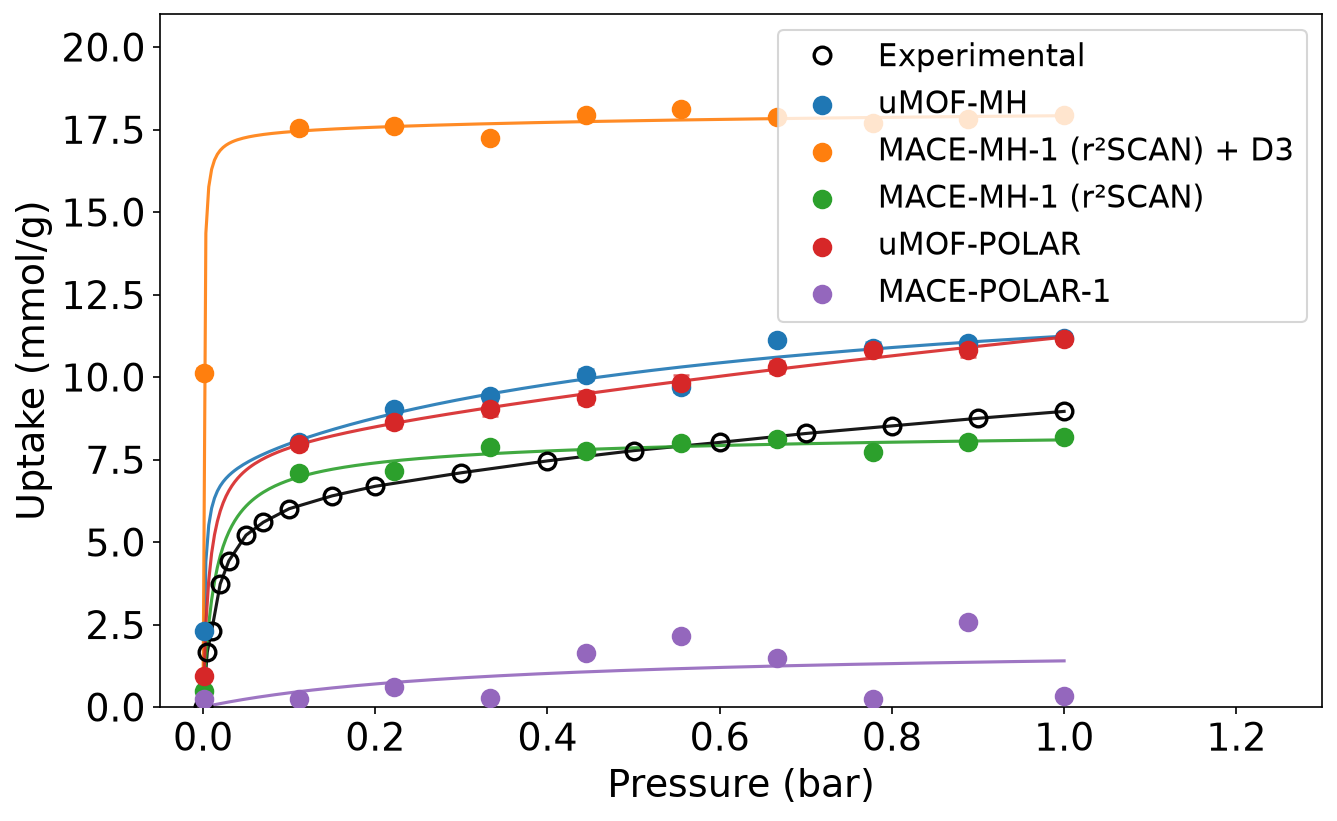}
\caption{Simulated CO$_2$ adsorption isotherms for Mg-MOF-74 at 298\,K (dual-site Langmuir fits to GCMC uptake) versus experiment. The uMOF models reproduce the experimental trajectory including the high-pressure plateau, whereas the dispersion-corrected foundation model overshoots and the uncorrected one saturates prematurely.}
\label{fig:isotherm}
\end{figure}

Beyond enthalpies, the shape of adsorption isotherm is a key observable, encoding the adsorption mechanism, pore-filling behavior and thermodynamic response across pressure that are critical for reticular chemistry. We performed rigid Grand Canonical Monte Carlo (GCMC) simulation on Mg-MOF-74 for CO$_2$ isotherms with each model. We found that adding a D3 correction to MACE-MH-1 overshoots the uptake (plateauing near 17.5\,mmol\,g$^{-1}$) through an overestimated adsorption enthalpy, whereas removing dispersion moves closer to experiment only through error compensation while still getting an incorrect shape. Our uMOF models, trained directly on r$^2$SCAN-D4 data with dispersion built in, instead reproduce the experimental isotherm shape with a slight plateau shift while the parent foundations saturate prematurely or are significantly off Figure~\ref{fig:isotherm}. The difference is a small systematic shift that additional CO$_2$-MOF training data should close. This agreement over the full isotherm shows that uMOF captures the broader adsorption physics required for reliable MOF screening.

\section{Conclusion}
\label{sec:conclusion}

uMOF pairs the largest and most accurate DFT dataset for MOF chemistry to date (85{,}524 r$^2$SCAN-D4 configurations, 19{,}950 frameworks, 79 elements) with an LLM-mined experimental benchmark (3{,}986 records, 20 property classes, distributed through the open \texttt{ml-peg} platform) and two fine-tuned MACE models, all released for community use. On generic near-equilibirum "Tier-1" properties the uMOF models match existing foundation and fine-tuned baselines. On gas adsorption, a substantially harder, dynamical task and a long-standing issue for MOF MLIPs, they show a clear, consistent advantage over every baseline tested, with predicted binding sites, geometries and isotherm that agree with experiment. This advantage traces to two complementary factors: a small but decisive fraction of finite-temperature MD trajectories that stabilizes MLIP out-of-equilibrium, and the r$^2$SCAN-D4 level of theory, which better captures MOF-guest dynamics than the classic PBE-based data underlying prior large-scale MOF datasets. uMOF's element-general representation should extend naturally to related reticular materials such as covalent-organic frameworks and polymers. Natural next step is broadening the benchmark's Tier-2 dynamical coverage and material classes beyond MOFs.

\section*{Data and Code Availability}
The uMOF dataset, benchmark, and model weights, together with training and evaluation code, are released publicly on figshare here: \url{https://doi.org/10.6084/m9.figshare.33311829}. The benchmark pipeline is distributed as a module of \texttt{ml-peg} at \url{https://github.com/ddmms/ml-peg}.
\section*{Acknowledgments}
This research used computing resources of Lawrencium, National Energy Research Scientific Computing Center (NERSC), a Department of Energy (DOE) Office of Science User Facility and by Science and Technology Facilities Council Scientific Computing Department’s SCARF cluster. T.J.I acknowledges financial support from the Bakar Institute of Digital Materials for the Planet. P.D.K. acknowledges financial support from U.S. National Science Foundation's "The Quantum Sensing Challenges for Transformational Advances in Quantum Systems" program No. 2326838. A.M.E. was also supported by the Ada Lovelace Centre at the Science and Technology Facilities Council (https://adalovelacecentre.ac.uk/), the Physical Sciences Data Infrastructure (https://psdi.ac.uk; jointly STFC and the University of Southampton) under grants EP/X032663/1 and EP/X032701/1, and EPSRC under grants EP/W026775/1 and EP/V028537/1.
\bibliographystyle{unsrtnat}
\bibliography{references,citation}

@article{insmof-5,
  title={Scrutinizing negative thermal expansion in MOF-5 by scattering techniques and ab initio calculations},
  author={Lock, Nina and Christensen, Mogens and Wu, Yue and Peterson, Vanessa K and Thomsen, Maja K and Piltz, Ross O and Ramirez-Cuesta, Anibal J and McIntyre, Garry J and Nor{\'e}n, Katarina and Kutteh, Ramzi and others},
  journal={Dalton Trans.},
  volume={42},
  number={6},
  pages={1996--2007},
  year={2013},
  publisher={Royal Society of Chemistry}
}

@article{Moosavi,
 title={A data-science approach to predict the heat capacity of nanoporous materials},
  author={Moosavi, Seyed Mohamad and Novotny, Bal{\'a}zs {\'A}lmos and Ongari, Daniele and Moubarak, Elias and Asgari, Mehrdad and Kadioglu, {\"O}zge and Charalambous, Charithea and Ortega-Guerrero, Andres and Farmahini, Amir H and Sarkisov, Lev and others},
  journal={Nat. Mater.},
  volume={21},
  number={12},
  pages={1419--1425},
  year={2022},
  publisher={Nature Publishing Group UK London}
}

@article{KLOUTSE20151,
title = {Specific heat capacities of MOF-5, Cu-BTC, Fe-BTC, MOF-177 and MIL-53 (Al) over wide temperature ranges: Measurements and application of empirical group contribution method},
journal = {Microporous Mesoporous Mater.},
volume = {217},
pages = {1-5},
year = {2015},
issn = {1387-1811},
doi = {https://doi.org/10.1016/j.micromeso.2015.05.047},
url = {https://www.sciencedirect.com/science/article/pii/S1387181115003248},
author = {F.A. Kloutse and R. Zacharia and D. Cossement and R. Chahine},

}

@article{10.1063/5.0201523,
    author = {Lai, Hoa Thi and Tran, Nhat Quang Minh and Nguyen, Linh Ho Thuy and Le, Thu Bao Nguyen and Nguyen, Cuong Chi and Pham, Anh Tuan Thanh and Doan, Tan Le Hoang and Park, Sungkyun and Hong, Jongill and Snyder, Gerald Jeffrey and Phan, Thang Bach},
    title = {Low experimental thermal conductivity of zirconium metal-organic framework UiO-66},
    journal = {Appl. Phys. Lett.},
    volume = {124},
    number = {15},
    pages = {152205},
    year = {2024},
    month = {04},
    issn = {0003-6951},
    doi = {10.1063/5.0201523},
    url = {https://doi.org/10.1063/5.0201523},
    eprint = {https://pubs.aip.org/aip/apl/article-pdf/doi/10.1063/5.0201523/19877867/152205_1_5.0201523.pdf},
}

@article{phonopy-phono3py-JPSJ,
  author  = {Togo, Atsushi},
  title   = {First-principles Phonon Calculations with Phonopy and Phono3py},
  journal = {J. Phys. Soc. Jpn.},
  volume  = {92},
  number  = {1},
  pages   = {012001},
  year    = {2023},
  doi     = {10.7566/JPSJ.92.012001}
}

@article{phonopy-phono3py-JPCM,
  author  = {Togo, Atsushi and Chaput, Laurent and Tadano, Terumasa and Tanaka, Isao},
  title   = {Implementation strategies in phonopy and phono3py},
  journal = {J. Phys. Condens. Matter},
  volume  = {35},
  number  = {35},
  pages   = {353001},
  year    = {2023},
  doi     = {10.1088/1361-648X/acd831}
}

@article{kamath,
  author  = {Elena, A. M. and Kamath, P. D. and Jaffrelot Inizan, T. and Rosen, A. S. and Zanca, F. and Persson, K. A.},
  title   = {Machine learned potential for high-throughput phonon calculations of metal--organic frameworks},
  year    = {2025},
  journal = {npj Comput. Mater.},
  volume  = {11},
  pages   = {125},
  doi     = {10.1038/s41524-025-01611-8}
}

@article{mofsim,
	author = {Kra{\ss}, Hendrik and Huang, Ju and Moosavi, Seyed Mohamad},
	date = {2025/12/11},
	doi = {10.1038/s41524-025-01872-3},
	id = {Kra{\ss}2025},
	isbn = {2057-3960},
	journal = {npj Comput. Mater.},
	number = {1},
	pages = {4},
	title = {MOFSimBench: evaluating universal machine learning interatomic potentials in metal-organic framework molecular modeling},
	url = {https://doi.org/10.1038/s41524-025-01872-3},
	volume = {12},
	year = {2025}}

@article{r2scanmof,
  author  = {Edzards, J. and Santana-Andreo, J. and Saßnick, H.-D. and Cocchi, C.},
  title   = {Benchmarking Selected Density Functionals and Dispersion Corrections for MOF-5 and Its Derivatives},
  year    = {2025},
  journal = {J. Chem. Theory Comput.},
  volume  = {21},
  pages   = {7062--7074},
  doi     = {10.1021/acs.jctc.5c00399}
}

@article{qmof1,
  author  = {Rosen, A. S. and Iyer, S. M. and Ray, D. and Yao, Z. and Aspuru-Guzik, A. and Gagliardi, L. and Notestein, J. M. and Snurr, R. Q.},
  title   = {Machine learning the quantum-chemical properties of metal--organic frameworks for accelerated materials discovery},
  year    = {2021},
  journal = {Matter},
  volume  = {4},
  pages   = {1578--1597},
  doi     = {10.1016/j.matt.2021.02.015}
}

@article{qmof2,
  author  = {Rosen, A. S. and Fung, V. and Huck, P. and O'Donnell, C. T. and Horton, M. K. and Truhlar, D. G. and Persson, K. A. and Notestein, J. M. and Snurr, R. Q.},
  title   = {High-throughput predictions of metal--organic framework electronic properties: theoretical challenges, graph neural networks, and data exploration},
  year    = {2022},
  journal = {npj Comput. Mater.},
  volume  = {8},
  pages   = {112},
  doi     = {10.1038/s41524-022-00796-6}
}

@article{mof,
  author  = {Zhou, H.-C. and Long, J. R. and Yaghi, O. M.},
  title   = {Introduction to Metal--Organic Frameworks},
  year    = {2012},
  journal = {Chem. Rev.},
  volume  = {112},
  pages   = {673--674},
  doi     = {10.1021/cr200018s}
}

@article{mof1,
author = {Hiroyasu Furukawa  and Kyle E. Cordova  and Michael O’Keeffe  and Omar M. Yaghi },
title = {The Chemistry and Applications of Metal-Organic Frameworks},
journal = {Science},
volume = {341},
number = {6149},
pages = {1230444},
year = {2013},
doi = {10.1126/science.1230444},
URL = {https://www.science.org/doi/abs/10.1126/science.1230444},
eprint = {https://www.science.org/doi/pdf/10.1126/science.1230444},
}

@article{mof2,
	author = {Yaghi, Omar M. and O'Keeffe, Michael and Ockwig, Nathan W. and Chae, Hee K. and Eddaoudi, Mohamed and Kim, Jaheon},
	date = {2003/06/01},
	doi = {10.1038/nature01650},
	id = {Yaghi2003},
	isbn = {1476-4687},
	journal = {Nature},
	number = {6941},
	pages = {705--714},
	title = {Reticular synthesis and the design of new materials},
	url = {https://doi.org/10.1038/nature01650},
	volume = {423},
	year = {2003}}

@article{vasp,
  title = {Efficient iterative schemes for ab initio total-energy calculations using a plane-wave basis set},
  author = {Kresse, G. and Furthm\"uller, J.},
  journal = {Phys. Rev. B},
  volume = {54},
  issue = {16},
  pages = {11169--11186},
  numpages = {0},
  year = {1996},
  month = {Oct},
  publisher = {American Physical Society},
  doi = {10.1103/PhysRevB.54.11169},
  url = {https://link.aps.org/doi/10.1103/PhysRevB.54.11169}
}

@article{vasp1,
  title = {From ultrasoft pseudopotentials to the projector augmented-wave method},
  author = {Kresse, G. and Joubert, D.},
  journal = {Phys. Rev. B},
  volume = {59},
  issue = {3},
  pages = {1758--1775},
  numpages = {0},
  year = {1999},
  month = {Jan},
  publisher = {American Physical Society},
  doi = {10.1103/PhysRevB.59.1758},
  url = {https://link.aps.org/doi/10.1103/PhysRevB.59.1758}
}

@article{vasp2,
title = {Efficiency of ab-initio total energy calculations for metals and semiconductors using a plane-wave basis set},
journal = {Comput. Mater. Sci.},
volume = {6},
number = {1},
pages = {15-50},
year = {1996},
issn = {0927-0256},
doi = {https://doi.org/10.1016/0927-0256(96)00008-0},
url = {https://www.sciencedirect.com/science/article/pii/0927025696000080},
author = {G. Kresse and J. Furthmüller}
}

@article{vasp3,
  title = {Ab initio molecular dynamics for liquid metals},
  author = {Kresse, G. and Hafner, J.},
  journal = {Phys. Rev. B},
  volume = {47},
  issue = {1},
  pages = {558--561},
  numpages = {0},
  year = {1993},
  month = {Jan},
  publisher = {American Physical Society},
  doi = {10.1103/PhysRevB.47.558},
  url = {https://link.aps.org/doi/10.1103/PhysRevB.47.558}
}

@article{d3,
    author = {Grimme, Stefan and Antony, Jens and Ehrlich, Stephan and Krieg, Helge},
    title = "{A consistent and accurate ab initio parametrization of density functional dispersion correction (DFT-D) for the 94 elements H-Pu}",
    journal = {J. Chem. Phys.},
    volume = {132},
    number = {15},
    pages = {154104},
    year = {2010},
    month = {04},
    issn = {0021-9606},
    doi = {10.1063/1.3382344},
    url = {https://doi.org/10.1063/1.3382344},
    eprint = {https://pubs.aip.org/aip/jcp/article-pdf/doi/10.1063/1.3382344/15684000/154104\_1\_online.pdf}
}

@article{bj,
author = {Grimme, Stefan and Ehrlich, Stephan and Goerigk, Lars},
title = {Effect of the damping function in dispersion corrected density functional theory},
journal = {J. Comput. Chem.},
volume = {32},
number = {7},
pages = {1456-1465},
doi = {https://doi.org/10.1002/jcc.21759},
url = {https://onlinelibrary.wiley.com/doi/abs/10.1002/jcc.21759},
eprint = {https://onlinelibrary.wiley.com/doi/pdf/10.1002/jcc.21759},
year = {2011}
}

@article{pbegga,
  title = {Generalized Gradient Approximation Made Simple},
  author = {Perdew, John P. and Burke, Kieron and Ernzerhof, Matthias},
  journal = {Phys. Rev. Lett.},
  volume = {77},
  issue = {18},
  pages = {3865--3868},
  numpages = {0},
  year = {1996},
  month = {Oct},
  publisher = {American Physical Society},
  doi = {10.1103/PhysRevLett.77.3865},
  url = {https://link.aps.org/doi/10.1103/PhysRevLett.77.3865}
}

@article{opendac,
	author = {Sriram A. and Choi S. and Yu X. and Brabson L. M. and Das A. and Ulissi Z. and Uyttendaele M. and Medford A. J. and Sholl D. S.},
    title = {The Open DAC 2023 Dataset and Challenges for Sorbent Discovery in Direct Air Capture},
    year    = {2024},
  journal = {ACS Cent. Sci.},
    volume = {10},
    pages = {923--941},
  doi = {10.1021/acscentsci.3c01629},
}

@article{r2scan,
  author  = {Furness J. W. and Kaplan A. D. and Ning J. and Perdew J. P. and Sun J.},
  title   = {Accurate and Numerically Efficient r2SCAN Meta-Generalized Gradient Approximation},
  year    = {2020},
  journal = {J. Phys. Chem. Lett.},
  volume  = {11},
  pages   = {8208--8215},
  doi     = {10.1021/acs.jpclett.0c02405}
}

@article{mptrj,
  author  = {Deng B. and Zhong P. and Jun K. and Riebesell J. and Han K. and Bartel C. J. and Ceder G.},
  title   = {CHGNet as a pretrained universal neural network potential for charge-informed atomistic modelling},
  year    = {2023},
  journal = {Nat. Mach. Intell.},
  volume  = {5},
  pages   = {1031--1041},
  doi     = {10.1038/s42256-023-00716-3}
}

@article{u,
  author  = {Rosen A. S. and Notestein J. M. and Snurr R. Q.},
  title   = {Comparing GGA, GGA+U, and meta-GGA functionals for redox-dependent binding at open metal sites in metal--organic frameworks},
  year    = {2020},
  journal = {J. Chem. Phys.},
  volume  = {152},
  pages   = {224101},
  doi     = {10.1063/5.0010166}
}

@software{janus-core,
  author       = {Kasoar, Elliott and
                  Austin, Patrick and
                  Devereux, Harvey and
                  Harris, Kieran and
                  Mason, David and
                  Wilkins, Jacob and
                  Zanca, Federica and
                  Elena, Alin},
  title        = {janus-core},
  license      = {BSD-3-Clause},
  repository-code = {https://github.com/stfc/janus-core},
  doi          = {10.5281/zenodo.13152495},
}

@inproceedings{batatia2022mace,
  title     = {{MACE}: Higher Order Equivariant Message Passing Neural Networks
               for Fast and Accurate Force Fields},
  author    = {Batatia, Ilyes and Kov{\'a}cs, D{\'a}vid P. and Simm, Gregor N. C.
               and Ortner, Christoph and Cs{\'a}nyi, G{\'a}bor},
  booktitle = {Advances in Neural Information Processing Systems (NeurIPS)},
  year      = {2022}
}

@article{chen2022universal,
  title={A universal graph deep learning interatomic potential for the periodic table},
  author={Chen, Chi and Ong, Shyue Ping},
  journal={Nature Computational Science},
  volume={2},
  number={11},
  pages={718--728},
  year={2022},
  publisher={Nature Publishing Group US New York}
}

@article{deng2023chgnet,
  title={CHGNet as a pretrained universal neural network potential for charge-informed atomistic modelling},
  author={Deng, Bowen and Zhong, Peichen and Jun, KyuJung and Riebesell, Janosh and Han, Kevin and Bartel, Christopher J and Ceder, Gerbrand},
  journal={Nature Machine Intelligence},
  volume={5},
  number={9},
  pages={1031--1041},
  year={2023},
  publisher={Nature Publishing Group UK London}
}

@article{barros2026open,
  title={The Open Materials 2024 (OMat24) inorganic materials dataset and models},
  author={Barros-Luque, Luis and Shuaibi, Muhammed and Fu, Xiang and Wood, Brandon M and Dzamba, Misko and Gao, Meng and Rizvi, Ammar and Uyttendaele, Matt and Zitnick, C Lawrence and Ulissi, Zachary W},
  journal={Nature Computational Science},
  pages={1--11},
  year={2026},
  publisher={Nature Publishing Group US New York}
}

@article{ghahremanpour2018alexandria,
  title={The Alexandria library, a quantum-chemical database of molecular properties for force field development},
  author={Ghahremanpour, Mohammad M and Van Maaren, Paul J and Van Der Spoel, David},
  journal={Scientific data},
  volume={5},
  number={1},
  pages={180062},
  year={2018},
  publisher={Nature Publishing Group}
}

@article{li2012metal,
  title={Metal--organic frameworks for separations},
  author={Li, Jian-Rong and Sculley, Julian and Zhou, Hong-Cai},
  journal={Chemical reviews},
  volume={112},
  number={2},
  pages={869--932},
  year={2012},
  publisher={ACS Publications}
}

@article{rehak2020including,
  title={Including dispersion in density functional theory for adsorption on flat oxide surfaces, in metal--organic frameworks and in acidic zeolites},
  author={Rehak, Florian R and Piccini, GiovanniMaria and Alessio, Maristella and Sauer, Joachim},
  journal={Physical Chemistry Chemical Physics},
  volume={22},
  number={14},
  pages={7577--7585},
  year={2020},
  publisher={The Royal Society of Chemistry}
}

@article{vlaisavljevich2017performance,
  title={Performance of van der Waals corrected functionals for guest adsorption in the M2 (dobdc) metal--organic frameworks},
  author={Vlaisavljevich, Bess and Huck, Johanna and Hulvey, Zeric and Lee, Kyuho and Mason, Jarad A and Neaton, Jeffrey B and Long, Jeffrey R and Brown, Craig M and Alfè, Dario and Michaelides, Angelos and others},
  journal={The journal of physical chemistry. A},
  volume={121},
  number={21},
  pages={4139},
  year={2017}
}

@article{cho2024improving,
  title={Improving gas adsorption modeling for MOFs by local calibration of Hubbard U parameters},
  author={Cho, Yeongsu and Kulik, Heather J},
  journal={The Journal of chemical physics},
  volume={160},
  number={15},
  year={2024},
  publisher={AIP Publishing}
}

@article{chiang2026mlip,
  title={MLIP arena: advancing fairness and transparency in machine learning interatomic potentials via an open, accessible benchmark platform},
  author={Chiang, Yuan and Kreiman, Tobias and Zhang, Christine and Kuner, Matthew and Weaver, Elizabeth and Amin, Ishan and Park, Hyunsoo and Lim, Yunsung and Kim, Jihan and Chrzan, Daryl and others},
  journal={Advances in Neural Information Processing Systems},
  volume={38},
  year={2026}
}

@article{sriram2025open,
  title={The Open DAC 2025 dataset for sorbent discovery in direct air capture},
  author={Sriram, Anuroop and Brabson, Logan M and Yu, Xiaohan and Choi, Sihoon and Abdelmaqsoud, Kareem and Moubarak, Elias and de Haan, Pim and L{\"o}we, Sindy and Brehmer, Johann and Kitchin, John R and others},
  journal={arXiv preprint arXiv:2508.03162},
  year={2025}
}

@article{lim2025accelerating,
  title={Accelerating CO2 direct air capture screening for metal-organic frameworks with a transferable machine learning force field},
  author={Lim, Yunsung and Park, Hyunsoo and Walsh, Aron and Kim, Jihan},
  journal={Matter},
  volume={8},
  number={7},
  year={2025},
  publisher={Elsevier}
}

@article{lee2009metal,
  title={Metal--organic framework materials as catalysts},
  author={Lee, JeongYong and Farha, Omar K and Roberts, John and Scheidt, Karl A and Nguyen, SonBinh T and Hupp, Joseph T},
  journal={Chemical Society Reviews},
  volume={38},
  number={5},
  pages={1450--1459},
  year={2009},
  publisher={The Royal Society of Chemistry}
}

@article{li2018recent,
  title={Recent advances in gas storage and separation using metal--organic frameworks},
  author={Li, Hao and Wang, Kecheng and Sun, Yujia and Lollar, Christina T and Li, Jialuo and Zhou, Hong-Cai},
  journal={Materials Today},
  volume={21},
  number={2},
  pages={108--121},
  year={2018},
  publisher={Elsevier}
}

@article{batatia2025foundation,
  title={A foundation model for atomistic materials chemistry},
  author={Batatia, Ilyes and Benner, Philipp and Chiang, Yuan and Elena, Alin M and Kov{\'a}cs, D{\'a}vid P and Riebesell, Janosh and Advincula, Xavier R and Asta, Mark and Avaylon, Matthew and Baldwin, William J and others},
  journal={The Journal of chemical physics},
  volume={163},
  number={18},
  year={2025},
  publisher={AIP Publishing}
}

@article{neumann2024orb,
  title={Orb: A fast, scalable neural network potential},
  author={Neumann, Mark and Gin, James and Rhodes, Benjamin and Bennett, Steven and Li, Zhiyi and Choubisa, Hitarth and Hussey, Arthur and Godwin, Jonathan},
  journal={arXiv preprint arXiv:2410.22570},
  year={2024}
}

@article{queen2014comprehensive,
  title={Comprehensive study of carbon dioxide adsorption in the metal--organic frameworks M2 (dobdc)(M= Mg, Mn, Fe, Co, Ni, Cu, Zn)},
  author={Queen, Wendy L and Hudson, Matthew R and Bloch, Eric D and Mason, Jarad A and Gonzalez, Miguel I and Lee, Jason S and Gygi, David and Howe, Joshua D and Lee, Kyuho and Darwish, Tamim A and others},
  journal={Chemical Science},
  volume={5},
  number={12},
  pages={4569--4581},
  year={2014},
  publisher={The Royal Society of Chemistry}
}

@article{lin2021scalable,
  title={A scalable metal-organic framework as a durable physisorbent for carbon dioxide capture},
  author={Lin, Jian-Bin and Nguyen, Tai TT and Vaidhyanathan, Ramanathan and Burner, Jake and Taylor, Jared M and Durekova, Hana and Akhtar, Farid and Mah, Roger K and Ghaffari-Nik, Omid and Marx, Stefan and others},
  journal={Science},
  volume={374},
  number={6574},
  pages={1464--1469},
  year={2021},
  publisher={American Association for the Advancement of Science}
}

@article{zhang2013enhancement,
  title={Enhancement of CO2 adsorption and CO2/N2 selectivity on ZIF-8 via postsynthetic modification},
  author={Zhang, Zhijuan and Xian, Shikai and Xia, Qibin and Wang, Haihui and Li, Zhong and Li, Jing},
  journal={AIChE Journal},
  volume={59},
  number={6},
  pages={2195--2206},
  year={2013},
  publisher={Wiley Online Library}
}

@article{abid2012nanosize,
  title={Nanosize Zr-metal organic framework (UiO-66) for hydrogen and carbon dioxide storage},
  author={Abid, Hussein Rasool and Tian, Huyong and Ang, Ha-Ming and Tade, Moses O and Buckley, Craig E and Wang, Shaobin},
  journal={Chemical Engineering Journal},
  volume={187},
  pages={415--420},
  year={2012},
  publisher={Elsevier}
}

@article{simmons2011carbon,
  title={Carbon capture in metal--organic frameworks—a comparative study},
  author={Simmons, Jason M and Wu, Hui and Zhou, Wei and Yildirim, Taner},
  journal={Energy \& Environmental Science},
  volume={4},
  number={6},
  pages={2177--2185},
  year={2011},
  publisher={The Royal Society of Chemistry}
}

@article{mishra2014adsorption,
  title={Adsorption and separation of carbon dioxide using MIL-53 (Al) metal-organic framework},
  author={Mishra, Prashant and Uppara, Hari Prasad and Mandal, Bishnupada and Gumma, Sasidhar},
  journal={Industrial \& Engineering Chemistry Research},
  volume={53},
  number={51},
  pages={19747--19753},
  year={2014},
  publisher={ACS Publications}
}

@article{mollmer2011high,
  title={High pressure adsorption of hydrogen, nitrogen, carbon dioxide and methane on the metal--organic framework HKUST-1},
  author={M{\"o}llmer, Jens and M{\"o}ller, Andreas and Dreisbach, Frieder and Gl{\"a}ser, Roger and Staudt, Reiner},
  journal={Microporous and Mesoporous Materials},
  volume={138},
  number={1-3},
  pages={140--148},
  year={2011},
  publisher={Elsevier}
}

@article{kamath2026data,
  title={Data-driven Design of Metal-Organic Frameworks with Tunable Negative Thermal Expansion},
  author={Kamath, Prathami Divakar and Tavani, Francesco and Elena, Alin Marin and Inizan, Th{\'e}o Jaffrelot and Lin, Yen-hsu and Yin, Jian and Xu, Wenqian and Yaghi, Omar M and Persson, Kristin A},
  journal={arXiv preprint arXiv:2607.18594},
  year={2026}
}

@article{vervoorts2019zeolitic,
  title={The Zeolitic Imidazolate Framework ZIF-4 under Low Hydrostatic Pressures},
  author={Vervoorts, Pia and Hobday, Claire L and Ehrenreich, Michael G and Daisenberger, Dominik and Kieslich, Gregor},
  journal={Zeitschrift f{\"u}r anorganische und allgemeine Chemie},
  volume={645},
  number={15},
  pages={970--974},
  year={2019},
  publisher={Wiley Online Library}
}

@article{kancharlapalli2023high,
  title={High-throughput screening of the CoRE-MOF-2019 database for CO2 capture from wet flue gas: a multi-scale modeling strategy},
  author={Kancharlapalli, Srinivasu and Snurr, Randall Q},
  journal={ACS Applied Materials \& Interfaces},
  volume={15},
  number={23},
  pages={28084--28092},
  year={2023},
  publisher={ACS Publications}
}

@article{ganose2025atomate2,
  title={Atomate2: Modular workflows for materials science},
  author={Ganose, Alex M and Sahasrabuddhe, Hrushikesh and Asta, Mark and Beck, Kevin and Biswas, Tathagata and Bonkowski, Alexander and Bustamante, Joana and Chen, Xin and Chiang, Yuan and Chrzan, Daryl C and others},
  journal={Digital discovery},
  volume={4},
  number={7},
  pages={1944--1973},
  year={2025},
  publisher={The Royal Society of Chemistry}
}

@article{wood2026family,
  title={UMA: A family of universal models for atoms},
  author={Wood, Brandon and Dzamba, Misko and Fu, Xiang and Gao, Meng and Shuaibi, Muhammed and Barroso-Luque, Luis and Abdelmaqsoud, Kareem and Gharakhanyan, Vahe and Kitchin, John and Levine, Daniel and others},
  journal={Advances in Neural Information Processing Systems},
  volume={38},
  pages={129391--129427},
  year={2026}
}

@article{edzards2025benchmarking,
  title={Benchmarking Selected Density Functionals and Dispersion Corrections for MOF-5 and Its Derivatives},
  author={Edzards, Joshua and Santana-Andreo, Julia and Sa{\ss}nick, Holger-Dietrich and Cocchi, Caterina},
  journal={Journal of Chemical Theory and Computation},
  volume={21},
  number={14},
  pages={7062},
  year={2025}
}

@article{chapman2008guest,
  title={Guest-dependent high pressure Phenomena in a nanoporous metal- organic framework material},
  author={Chapman, Karena W and Halder, Gregory J and Chupas, Peter J},
  journal={Journal of the American Chemical Society},
  volume={130},
  number={32},
  pages={10524--10526},
  year={2008},
  publisher={ACS Publications}
}

@article{yot2016exploration,
  title={Exploration of the mechanical behavior of metal organic frameworks UiO-66 (Zr) and MIL-125 (Ti) and their NH2 functionalized versions},
  author={Yot, Pascal G and Yang, Ke and Ragon, Florence and Dmitriev, Vladimir and Devic, Thomas and Horcajada, Patricia and Serre, Christian and Maurin, Guillaume},
  journal={Dalton Transactions},
  volume={45},
  number={10},
  pages={4283--4288},
  year={2016},
  publisher={The Royal Society of Chemistry}
}

@article{yot2014metal,
  title={Metal--organic frameworks as potential shock absorbers: the case of the highly flexible MIL-53 (Al)},
  author={Yot, Pascal G and Boudene, Zoubeyr and Macia, Jasmine and Granier, Dominique and Vanduyfhuys, Louis and Verstraelen, Toon and Van Speybroeck, Veronique and Devic, Thomas and Serre, Christian and F{\'e}rey, G{\'e}rard and others},
  journal={Chemical Communications},
  volume={50},
  number={67},
  pages={9462--9464},
  year={2014},
  publisher={The Royal Society of Chemistry}
}

@article{redfern2020isolating,
  title={Isolating the role of the node-linker bond in the compression of UiO-66 metal--organic frameworks},
  author={Redfern, Louis R and Ducamp, Maxime and Wasson, Megan C and Robison, Lee and Son, Florencia A and Coudert, Fran{\c{c}}ois-Xavier and Farha, Omar K},
  journal={Chemistry of Materials},
  volume={32},
  number={13},
  pages={5864--5871},
  year={2020},
  publisher={ACS Publications}
}

@article{zheng2017theoretical,
  title={Theoretical prediction of the mechanical properties of zeolitic imidazolate frameworks (ZIFs)},
  author={Zheng, Bin and Zhu, Yihan and Fu, Fang and Wang, Lian Li and Wang, Jinlei and Du, Huiling},
  journal={RSC advances},
  volume={7},
  number={66},
  pages={41499--41503},
  year={2017},
  publisher={The Royal Society of Chemistry}
}

@article{ryder2017detecting,
  title={Detecting molecular rotational dynamics complementing the low-frequency terahertz vibrations in a zirconium-based metal-organic framework},
  author={Ryder, Matthew R and Van de Voorde, Ben and Civalleri, Bartolomeo and Bennett, Thomas D and Mukhopadhyay, Sanghamitra and Cinque, Gianfelice and Fernandez-Alonso, Felix and De Vos, Dirk and Rudi{\'c}, Svemir and Tan, Jin-Chong},
  journal={Physical Review Letters},
  volume={118},
  number={25},
  pages={255502},
  year={2017},
  publisher={APS}
}

@article{ryder2014identifying,
  title={Identifying the role of terahertz vibrations in metal-organic frameworks: from gate-opening phenomenon to shear-driven structural destabilization},
  author={Ryder, Matthew R and Civalleri, Bartolomeo and Bennett, Thomas D and Henke, Sebastian and Rudi{\'c}, Svemir and Cinque, Gianfelice and Fernandez-Alonso, Felix and Tan, Jin-Chong},
  journal={Physical review letters},
  volume={113},
  number={21},
  pages={215502},
  year={2014},
  publisher={APS}
}

@article{liu2008reversible,
  title={Reversible structural transition in MIL-53 with large temperature hysteresis},
  author={Liu, Yun and Her, Jae-Hyuk and Dailly, Anne and Ramirez-Cuesta, Anibal J and Neumann, Dan A and Brown, Craig M},
  journal={Journal of the American Chemical Society},
  volume={130},
  number={35},
  pages={11813--11818},
  year={2008},
  publisher={ACS Publications}
}

@article{grajciar2011understanding,
  title={Understanding CO2 adsorption in CuBTC MOF: comparing combined DFT--ab initio calculations with microcalorimetry experiments},
  author={Grajciar, Lukas and Wiersum, Andrew D and Llewellyn, Philip L and Chang, Jong-San and Nachtigall, Petr},
  journal={The Journal of Physical Chemistry C},
  volume={115},
  number={36},
  pages={17925--17933},
  year={2011},
  publisher={ACS Publications}
}

@article{wasik2024impact,
  title={The impact of metal centers in the M-MOF-74 series on carbon dioxide and hydrogen separation},
  author={Wasik, Dominika O and Vicent-Luna, Jos{\'e} Manuel and Luna-Triguero, Azahara and Dubbeldam, David and Vlugt, Thijs JH and Calero, Sof{\'\i}a},
  journal={Separation and Purification Technology},
  volume={339},
  pages={126539},
  year={2024},
  publisher={Elsevier}
}

@article{oliveira2026flames,
  title={FLAMES--A flexible and extensible code for Monte Carlo simulations of nanoporous materials},
  author={Oliveira, Felipe L and Sa{\ss}nick, Holger-Dietrich and Maurin, Guillaume},
  year={2026}
}

@article{batatia2024foundation,
  title   = {A foundation model for atomistic materials chemistry},
  author  = {Batatia, Ilyes and Benner, Philipp and Chiang, Yuan and others},
  journal = {arXiv preprint arXiv:2401.00096},
  year    = {2024}
}

@article{furness2020r2scan,
  title   = {Accurate and Numerically Efficient {r$^2$SCAN} Meta-Generalized
             Gradient Approximation},
  author  = {Furness, James W. and Kaplan, Aaron D. and Ning, Jinliang and
             Perdew, John P. and Sun, Jianwei},
  journal = {The Journal of Physical Chemistry Letters},
  volume  = {11},
  number  = {19},
  pages   = {8208--8215},
  year    = {2020}
}

@article{caldeweyher2019d4,
  title   = {A generally applicable atomic-charge dependent {London} dispersion
             correction},
  author  = {Caldeweyher, Eike and Ehlert, Sebastian and Hansen, Andreas and
             Neugebauer, Hagen and Spicher, Sebastian and Bannwarth, Christoph
             and Grimme, Stefan},
  journal = {The Journal of Chemical Physics},
  volume  = {150},
  number  = {15},
  pages   = {154122},
  year    = {2019}
}

@article{rosen2021qmof,
  title   = {Machine learning the quantum-chemical properties of metal--organic
             frameworks for accelerated materials discovery},
  author  = {Rosen, Andrew S. and Iyer, Shaelyn M. and Ray, Debmalya and
             Yao, Zhenpeng and Aspuru-Guzik, Al{\'a}n and Gagliardi, Laura and
             Notestein, Justin M. and Snurr, Randall Q.},
  journal = {Matter},
  volume  = {4},
  number  = {5},
  pages   = {1578--1597},
  year    = {2021}
}

@article{chung2019coremof,
  title   = {Advances, Updates, and Analytics for the Computation-Ready,
             Experimental Metal--Organic Framework Database: {CoRE MOF 2019}},
  author  = {Chung, Yongchul G. and others},
  journal = {Journal of Chemical \& Engineering Data},
  volume  = {64},
  number  = {12},
  pages   = {5985--5998},
  year    = {2019}
}

@article{grisafi2019longrange,
  title   = {Incorporating long-range physics in atomic-scale machine learning},
  author  = {Grisafi, Andrea and Ceriotti, Michele},
  journal = {The Journal of Chemical Physics},
  volume  = {151},
  number  = {20},
  pages   = {204105},
  year    = {2019}
}

@article{merchant2023gnome,
  title   = {Scaling deep learning for materials discovery},
  author  = {Merchant, Amil and Batzner, Simon and Schoenholz, Samuel S. and
             Aykol, Muratahan and Cheon, Gowoon and Cubuk, Ekin Dogus},
  journal = {Nature},
  volume  = {624},
  pages   = {80--85},
  year    = {2023}
}

@article{riebesell2024matbench,
  title   = {Matbench Discovery: A framework to evaluate machine learning
             crystal stability predictions},
  author  = {Riebesell, Janosh and Goodall, Rhys E. A. and Benner, Philipp and
             others},
  journal = {arXiv preprint arXiv:2308.14920},
  year    = {2024}
}

@article{kaplan2025matpes,
  title   = {{MatPES}: A foundational potential energy surface dataset for
             materials},
  author  = {Kaplan, Aaron D. and others},
  journal = {arXiv preprint arXiv:2503.04070},
  year    = {2025}
}

@article{kamath2026impact,
  title={The impact of spurious imaginary phonon modes on thermal properties of Metal-organic Frameworks},
  author={Kamath, Prathami Divakar and Persson, Kristin A},
  journal={npj Computational Materials},
  year={2026},
  publisher={Nature Publishing Group UK London}
}

@article{park2017reproducible,
  title={How reproducible are isotherm measurements in metal--organic frameworks?},
  author={Park, Jongwoo and Howe, Joshua D and Sholl, David S},
  journal={Chemistry of Materials},
  volume={29},
  number={24},
  pages={10487--10495},
  year={2017},
  publisher={ACS Publications}
}

@article{longrange,
title = {Pore opening and breathing transitions in metal-organic frameworks: Coupling adsorption and deformation},
journal = {Journal of Colloid and Interface Science},
volume = {578},
pages = {77-88},
year = {2020},
issn = {0021-9797},
doi = {https://doi.org/10.1016/j.jcis.2020.05.105},
url = {https://www.sciencedirect.com/science/article/pii/S0021979720307165},
author = {Filip Formalik and Alexander V. Neimark and Justyna Rogacka and Lucyna Firlej and Bogdan Kuchta}
}

@article{caldeweyher2019generally,
  title={A generally applicable atomic-charge dependent London dispersion correction},
  author={Caldeweyher, Eike and Ehlert, Sebastian and Hansen, Andreas and Neugebauer, Hagen and Spicher, Sebastian and Bannwarth, Christoph and Grimme, Stefan},
  journal={The Journal of chemical physics},
  volume={150},
  number={15},
  year={2019},
  publisher={AIP Publishing}
}

@article{kang2025harnessing,
  title={Harnessing large language models to collect and analyze metal--organic framework property data set},
  author={Kang, Yeonghun and Lee, Wonseok and Bae, Taeun and Han, Seunghee and Jang, Huiwon and Kim, Jihan},
  journal={Journal of the American Chemical Society},
  volume={147},
  number={5},
  pages={3943--3958},
  year={2025},
  publisher={ACS Publications}
}

@article{eckhoff2019mof5,
  title   = {From Molecular Fragments to the Bulk: Development of a Neural
             Network Potential for {MOF-5}},
  author  = {Eckhoff, Marco and Behler, J{\"o}rg},
  journal = {Journal of Chemical Theory and Computation},
  volume  = {15},
  number  = {6},
  pages   = {3793--3809},
  year    = {2019},
  doi     = {10.1021/acs.jctc.8b01288}
}

@article{vandenhaute2023incremental,
  title   = {Machine learning potentials for metal--organic frameworks using an
             incremental learning approach},
  author  = {Vandenhaute, Sander and Cools-Ceuppens, Maarten and DeKeyser, Simon
             and Verstraelen, Toon and Van Speybroeck, Veronique},
  journal = {npj Computational Materials},
  volume  = {9},
  number  = {1},
  pages   = {19},
  year    = {2023},
  doi     = {10.1038/s41524-023-00969-x}
}

@article{wieser2024mlff,
  title   = {Machine learned force-fields for an ab-initio quality description of
             metal--organic frameworks},
  author  = {Wieser, Sandro and Zojer, Egbert},
  journal = {npj Computational Materials},
  volume  = {10},
  number  = {1},
  pages   = {18},
  year    = {2024},
  doi     = {10.1038/s41524-024-01205-w}
}

@article{sharma2024tempactive,
  title   = {Quantum-accurate machine learning potentials for metal--organic
             frameworks using temperature driven active learning},
  author  = {Sharma, Ankit and Sanvito, Stefano},
  journal = {npj Computational Materials},
  volume  = {10},
  number  = {1},
  pages   = {237},
  year    = {2024},
  doi     = {10.1038/s41524-024-01427-y}
}

@article{goeminne2023dftquality,
  title   = {{DFT}-Quality Adsorption Simulations in Metal--Organic Frameworks
             Enabled by Machine Learning Potentials},
  author  = {Goeminne, Ruben and Vanduyfhuys, Louis and Van Speybroeck, Veronique
             and Verstraelen, Toon},
  journal = {Journal of Chemical Theory and Computation},
  volume  = {19},
  number  = {18},
  pages   = {6313--6325},
  year    = {2023},
  doi     = {10.1021/acs.jctc.3c00495}
}

@article{liu2024h2oms,
  title   = {Machine learning potential for modelling {H$_2$}
             adsorption/diffusion in {MOFs} with open metal sites},
  author  = {Liu, Shanping and Dupuis, Romain and Fan, Dong and Benzaria, Salma
             and Bonneau, Mickaele and Bhatt, Prashant and Eddaoudi, Mohamed
             and Maurin, Guillaume},
  journal = {Chemical Science},
  volume  = {15},
  number  = {14},
  pages   = {5294--5302},
  year    = {2024},
  doi     = {10.1039/d3sc05612k}
}

@article{sriram2024odac,
  title   = {The Open {DAC} 2023 Dataset and Challenges for Sorbent Discovery in
             Direct Air Capture},
  author  = {Sriram, Anuroop and Choi, Sihoon and Yu, Xiaohan and Brabson,
             Logan M. and Das, Abhishek and Ulissi, Zachary and Uyttendaele, Matt
             and Medford, Andrew J. and Sholl, David S.},
  journal = {ACS Central Science},
  volume  = {10},
  number  = {5},
  pages   = {923--941},
  year    = {2024},
  doi     = {10.1021/acscentsci.3c01629}
}

@article{kuner2025mp,
  title={MP-ALOE: an r2SCAN dataset for universal machine learning interatomic potentials},
  author={Kuner, Matthew C and Kaplan, Aaron D and Persson, Kristin A and Asta, Mark and Chrzan, Daryl C},
  journal={npj Computational Materials},
  volume={11},
  number={1},
  pages={352},
  year={2025},
  publisher={Nature Publishing Group UK London}
}

@article{gibaldi2025mosaec,
  title={MOSAEC-DB: a comprehensive database of experimental metal--organic frameworks with verified chemical accuracy suitable for molecular simulations},
  author={Gibaldi, Marco and Kapeliukha, Anna and White, Andrew and Luo, Jun and Mayo, Robert Alex and Burner, Jake and Woo, Tom K},
  journal={Chemical Science},
  volume={16},
  number={9},
  pages={4085--4100},
  year={2025},
  publisher={The Royal Society of Chemistry}
}

@misc{mlpeg,
  title        = {{ml-peg}: ML Performance and Extrapolation Guide},
  author       = {{DDMMS, STFC}},
  howpublished = {\url{https://github.com/ddmms/ml-peg}},
  note         = {Reference deployment: \url{https://ml-peg.stfc.ac.uk}},
  year         = {2025}
}

@misc{inizan2025agenticaidiscoverymetalorganic,
      title={System of Agentic AI for the Discovery of Metal-Organic Frameworks}, 
      author={Theo Jaffrelot Inizan and Sherry Yang and Aaron Kaplan and Yen-hsu Lin and Jian Yin and Saber Mirzaei and Mona Abdelgaid and Ali H. Alawadhi and KwangHwan Cho and Zhiling Zheng and Ekin Dogus Cubuk and Christian Borgs and Jennifer T. Chayes and Kristin A. Persson and Omar M. Yaghi},
      year={2025},
      eprint={2504.14110},
      archivePrefix={arXiv},
      primaryClass={cond-mat.mtrl-sci},
      url={https://arxiv.org/abs/2504.14110}, 
}

@article{ehlert2021r2scand4,
  title   = {r$^2$SCAN-D4: Dispersion corrected meta-generalized gradient approximation for general chemical applications},
  author  = {Ehlert, Sebastian and Huniar, Uwe and Ning, Jinliang and Furness, James W. and Sun, Jianwei and Kaplan, Aaron D. and Perdew, John P. and Brandenburg, Jan Gerit},
  journal = {The Journal of Chemical Physics},
  volume  = {154}, number = {6}, pages = {061101}, year = {2021},
  doi     = {10.1063/5.0041008}
}

@article{dohm2018mor41,
  title   = {Comprehensive Thermochemical Benchmark Set of Realistic Closed-Shell Metal Organic Reactions},
  author  = {Dohm, Sebastian and Hansen, Andreas and Steinmetz, Marc and Grimme, Stefan and Checinski, Marek P.},
  journal = {Journal of Chemical Theory and Computation},
  volume  = {14}, number = {5}, pages = {2596--2608}, year = {2018},
  doi     = {10.1021/acs.jctc.7b01183}
}

\clearpage
\appendix

\section{DFT functional benchmarking against experiment}
\label{sec:appendix_dft}
\begin{figure}[h]
    \centering
    \includegraphics[width=\linewidth]{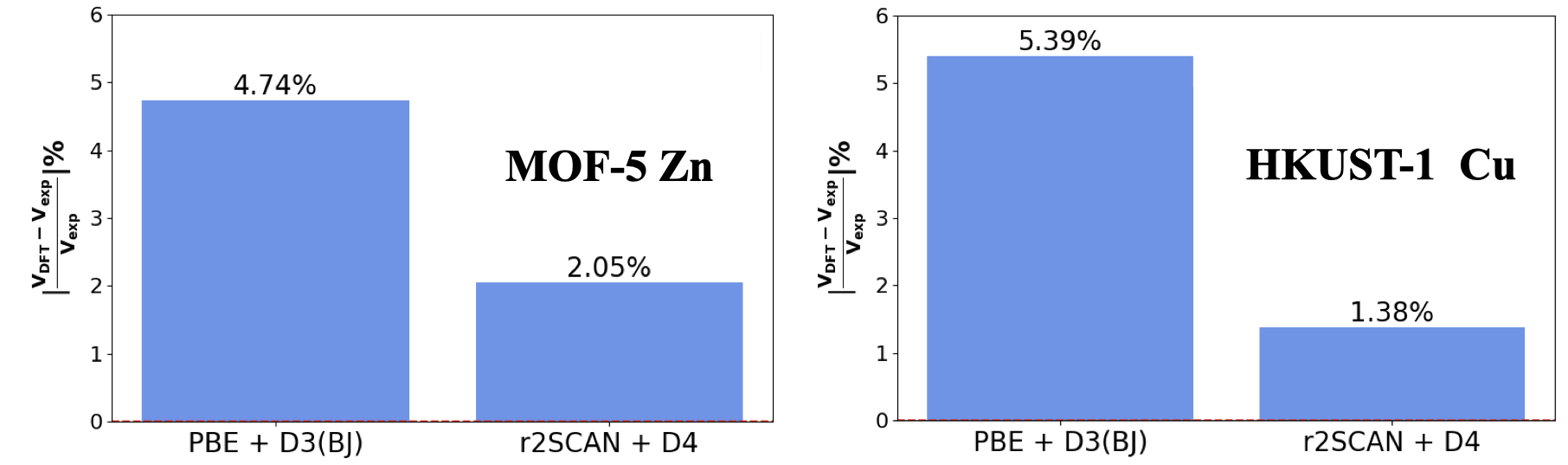}
    \caption{\textbf{Benchmarking DFT functionals.} r$^2$SCAN-D4 produces 2--4\% lower errors than PBE-D3(BJ) relative to experiment for optimized geometries of Zn-MOF-5 and Cu-HKUST-1.}
    \label{fig:r2scan}
\end{figure}

Since current large-scale MOF MLIP training data is overwhelmingly built at the PBE+D3(BJ) level, we benchmarked DFT-optimized geometries with r$^2$SCAN-D4 and PBE-D3(BJ) relative to experiment for two representative MOFs: Zn-MOF-5 (non-spin-polarized) and Cu-HKUST-1 (spin-polarized). r$^2$SCAN-D4 produces 2--4\% lower geometry errors than PBE-D3(BJ) across both frameworks (Figure~\ref{fig:r2scan}), with the improvement most pronounced for spin-polarized Cu-HKUST-1, where uncorrected PBE's larger delocalization error leads to overestimated cell volumes relative to experiment \citep{r2scan}. The trade-off is a modest overestimation of bulk modulus and a blue-shift in phonon frequencies \citep{edzards2025benchmarking}, consistent with the slightly higher bulk-modulus errors and blue-shifted INS spectra observed for r$^2$SCAN-D4-based MLIPs in Table~\ref{tab:tier1} and Figure~\ref{fig:ins}.

\section{Benchmark structure selection}
\label{sec:bench}
\paragraph{Bulk modulus.} Nine MOFs spanning seven unique metal nodes, with experimental references: Cu-HKUST-1 \citep{chapman2008guest}, Ti-MIL-125 \citep{yot2016exploration}, Al-MIL-53 \citep{yot2014metal}, Zr-UiO-66, Hf-UiO-66, and Ce-UiO-66 \citep{redfern2020isolating}, Zr-UiO-67, Zn-ZIF-8 \citep{zheng2017theoretical}, and Zn-ZIF-4 \citep{vervoorts2019zeolitic}.

\paragraph{Heat capacity.} Seven MOFs, with experimental references: Zn-MOF-5, Cu-MOF-177, Cu-HKUST-1, and Al-MIL-53 \citep{KLOUTSE20151}, Zr-UiO-66 \citep{10.1063/5.0201523}, and Zn-MOF-74 and Zn-ZIF-8 \citep{Moosavi}.

\begin{figure*}[t]
    \centering
    \begin{subfigure}{0.3\textwidth}
        \centering
        \includegraphics[width=\linewidth]{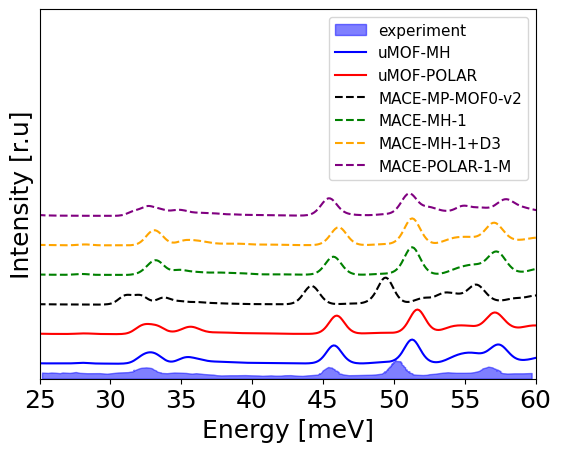}
        \caption{}
        \label{fig:ins1}
    \end{subfigure}
    \hfill
    \begin{subfigure}{0.3\textwidth}
        \centering
        \includegraphics[width=\linewidth]{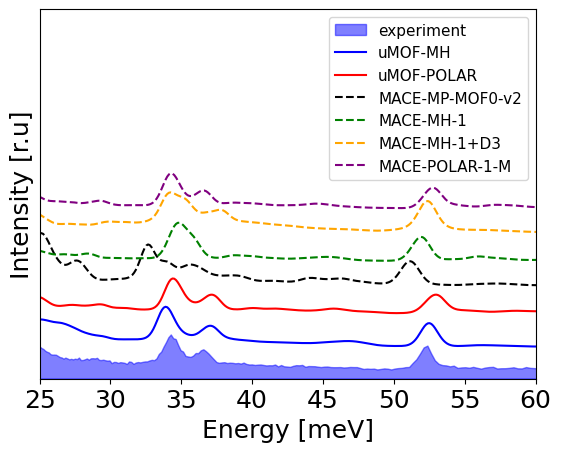}
        \caption{}
        \label{fig:ins2}
    \end{subfigure}
    \hfill
    \begin{subfigure}{0.3\textwidth}
        \centering
        \includegraphics[width=\linewidth]{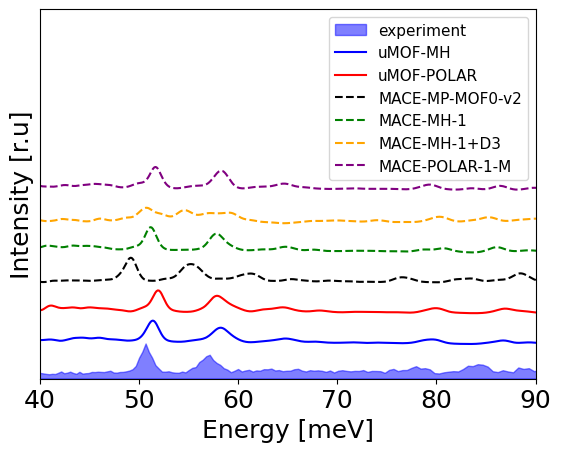}
        \caption{}
        \label{fig:ins3}
    \end{subfigure}
    \\[0.5em]
    \begin{subfigure}{0.3\textwidth}
        \centering
        \includegraphics[width=\linewidth]{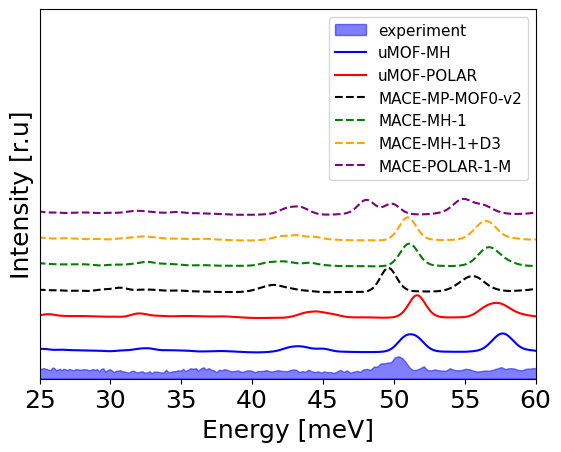}
        \caption{}
        \label{fig:ins4}
    \end{subfigure}
    \hfill
    \begin{subfigure}{0.3\textwidth}
        \centering
        \includegraphics[width=\linewidth]{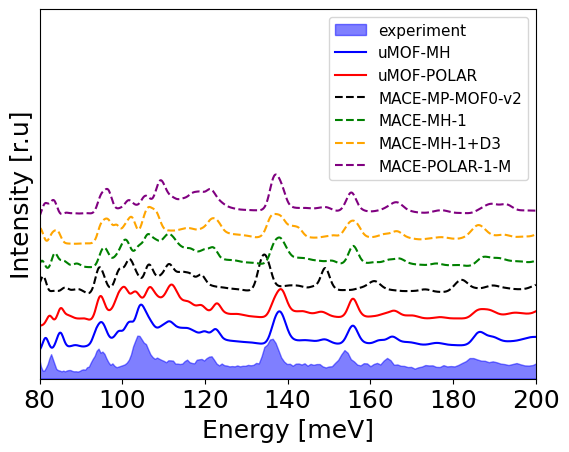}
        \caption{}
        \label{fig:ins5}
    \end{subfigure}
    \hfill
    \begin{subfigure}{0.3\textwidth}
        \centering
        \includegraphics[width=\linewidth]{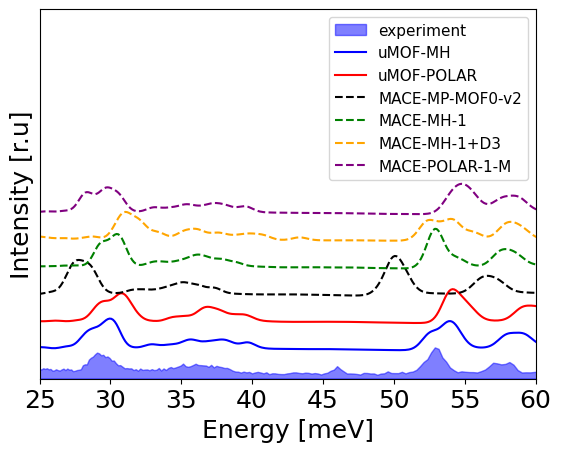}
        \caption{}
        \label{fig:ins6}
    \end{subfigure}
    \caption{Predicted vs.\ experimental inelastic neutron scattering (INS) spectra for six representative frameworks: (a) MOF-5, (b) ZIF-8, (c) MIL-53, (d) MIL-140A, (e) ZIF-4, (f) ZIF-7. Experimental data from \citet{insmof-5}, \citet{liu2008reversible}, \citet{ryder2014identifying}, and \citet{ryder2017detecting}.}
    \label{fig:ins}
\end{figure*}

\paragraph{Inelastic neutron scattering (INS).} Six MOFs, with experimental references: Zn-MOF-5 \citep{insmof-5}, Al-MIL-53 \citep{liu2008reversible}, Zn-ZIF-8, Zn-ZIF-4, and Zn-ZIF-7 \citep{ryder2014identifying}, and Zr-MIL-140A \citep{ryder2017detecting}. The ZIF series primarily differ in the functional groups and side chains attached to the base imidazole linker, providing a sensitive probe of model accuracy for the effect of linker functionalization on framework dynamics (Figure~\ref{fig:ins}).

\paragraph{Widom Insertion.} 13 MOFs, with experimental references: Zn-CALF-20 \citep{lin2021scalable}, Cu-HKUST-1\cite{mollmer2011high}, Zn-MOF-5 \citep{simmons2011carbon}, Al-MIL-53 \citep{mishra2014adsorption}, Zn-ZIF-8 \citep{zhang2013enhancement}, Zr-UiO-66 \citep{abid2012nanosize}, and the seven M-MOF-74 series \citep{queen2014comprehensive} for CO$_2$ adsorption. With three additional points where HKUST-1 was used for methane, nitrogen and hydrogen adsorption calculations. \cite{mollmer2011high}.

\section{Calculation parameters for benchmark evaluation}
\label{sec:appendix_params}
\paragraph{Bulk modulus.} Full-cell LBFGS optimization to fmax~$=10^{-6}$\,eV/\AA{}, followed by fixed-cell optimization at the same convergence for six strained configurations ($\pm3\%$ linear volumetric strain) using the \texttt{janus} package \citep{janus-core}; bulk modulus is obtained from a seven-point Birch--Murnaghan equation-of-state fit about the relaxed volume.

\paragraph{Heat capacity.} Harmonic phonons computed from 0 to 1000\,K on an $11\times11\times11$ mesh with Phonopy \citep{phonopy-phono3py-JPCM, phonopy-phono3py-JPSJ}, following full-cell LBFGS optimization to fmax~$=10^{-8}$\,eV/\AA{} or 1000 steps, whichever is reached first.

\paragraph{INS spectra.} Computed using the workflow of \citet{kamath2026data} with matching settings.

\paragraph{Enthalpy of adsorption.} Widom-insertion calculations performed with the \texttt{flames} package \citep{oliveira2026flames} at 298\,K for 20{,}000 steps; convergence with respect to step count was checked on HKUST-1 as a representative example, with 20{,}000 steps selected as balancing cost and accuracy.

\end{document}